\documentclass[reprint,twocolumn,groupedaddress,showkeys,floatfix,superscriptaddress,aps]{revtex4-2}
\usepackage[colorlinks=true,citecolor=red,filecolor=green,linkcolor=blue,pdfnewwindow=true]{hyperref}
\usepackage{amsmath} 
\usepackage[version=4]{mhchem}
\usepackage{graphicx}
\usepackage{bm}
\usepackage[title]{appendix} 
\usepackage{float}
\usepackage[normalem]{ulem}

\usepackage{setspace}

\usepackage{xcolor}
\usepackage{hyperref}
\usepackage{natbib}
\usepackage{footnote}

\hypersetup{urlcolor=blue}

\makeatletter
 
\newcommand{\Rmnum}[1]{\expandafter\@slowromancap\romannumeral #1@}

\usepackage{atbegshi}
\newcommand\showtimer{%
  \message{^^Jtimer: \the\numexpr\the\pdfelapsedtime*1000/65536\relax}%
  \pdfresettimer}
\AtBeginDocument{\showtimer}
\AtBeginShipout {\showtimer}

\makeatother
\begin{document}

\title{Exploring the Interplay of Intrinsic Fluctuation and Complexity in \\
Intracellular Calcium Dynamics}

\author{Athokpam Langlen Chanu}                        
\email{athokpam.chanu@apctp.org (Corresponding author)}
\affiliation{Asia Pacific Center for Theoretical Physics, Pohang, 37673, Republic of Korea}

\author{R.~K.~Brojen Singh}
\email{brojen@jnu.ac.in}
\affiliation{School of Computational and Integrative Sciences, Jawaharlal Nehru~University, New Delhi, 110067, India}

\author{Jae-Hyung Jeon}
\email{jeonjh@postech.ac.kr}
\affiliation{Asia Pacific Center for Theoretical Physics, Pohang, 37673, Republic of Korea}
\affiliation{Department of Physics, Pohang University of Science and Technology (POSTECH), Pohang, 37673, Republic of Korea}


\begin{abstract}
The concentration of intracellular calcium ion (Ca$^{2+}$) exhibits complex oscillations, including bursting and chaos, as observed experimentally. 
These dynamics are influenced by inherent fluctuations within cells, which serve as crucial determinants in cellular decision-making processes and fate determination. In this study, we systematically explore the interplay between intrinsic fluctuation and the complexity of various dynamic states of intracellular cytosolic Ca$^{2+}$. To investigate this interplay, we employ complexity measures such as permutation entropy and statistical complexity. Using a stochastic chemical Langevin equation, we simulate the dynamics of cytosolic Ca$^{2+}$. Our findings reveal that permutation entropy and statistical complexity effectively characterize the diverse, dynamic states of cytosolic Ca$^{2+}$ and illustrate their interactions with intrinsic fluctuation. Permutation entropy analysis elucidates that the chaotic
state is more sensitive to intrinsic fluctuation than the
other periodic states. Furthermore, we identify distinct states of cytosolic Ca$^{2+}$ occupying specific locations within the theoretical bounds of the complexity-entropy causality plane. These locations indicate varying complexity and information content as intrinsic fluctuation varies. When adjusting the permutation order, the statistical complexity measure for the different periodic and chaotic states exhibits peaks in an intermediate range of intrinsic fluctuation values. Additionally, we identify scale-free or fractal patterns in this intermediate range, which are further corroborated by multifractal detrended fluctuation analysis. These high-complexity states likely correspond to optimal Ca$^{2+}$ dynamics with biological significance, revealing rich and complex dynamics shaped by the interplay of intrinsic fluctuation and complexity. Our investigation enhances our understanding of the intricate regulatory mechanisms governing intracellular Ca$^{2+}$ dynamics and how intrinsic fluctuation modulates the complexity of the various dynamics that play crucial roles in biological cells.
\end{abstract}


\maketitle

\section{Introduction}
Complex systems comprise numerous interacting sub-units or components, characterized by intricate mutual interactions~\cite{mitchell2009complexity}. These interactions give rise to non-linear phenomena, including bifurcation~\cite{strogartz1994nonlinear}, which is considered an elementary act of complexity~\cite{nicolis1989exploring,nicolis2012foundations}. Other intriguing non-linear phenomena encompass multistability~\cite{pisarchik2014control}, chaos~\cite{rickles2007simple}, and fractals~\cite{aguirre2009fractal}. At microscopic and mesoscopic scales, diverse biological systems showcase fluctuation-driven stochastic dynamics, observed in biological cells~\cite{bressloff2014stochastic}, neural firing~\cite{allen1994evaluation}, biochemical networks~\cite{shahrezaei2008stochastic}, genetic oscillator networks~\cite{blossey2006compositional}, molecular motors~\cite{kolomeisky2013motor}, and more. Biological cells, complex biochemical systems~\cite{qian2007phosphorylation,qian2013stochastic}, exchange energy and matter with the surrounding environment, operating far from thermodynamic equilibrium. They exhibit dissipative structures~\cite{prigogine1978time}, such as temporal oscillations, waves, and patterns~\cite{epstein1998introduction,goldbeter2018dissipative,chung2022thermodynamics}, which exhibit a blend of orderliness (coherence) and randomness~\cite{poon1995controlling}. Exploring the complexity of dissipative structures like biochemical oscillations is intriguing and fundamentally significant.   

In complex biochemical systems like living cells, fluctuations manifest as intrinsic and extrinsic fluctuations~\cite{hilfinger2011separating}. Extrinsic fluctuations result from external factors, whereas intrinsic fluctuations arise from inherent random molecular interactions within a chemically reacting system~\cite{rao2002control}. These interactions correspond to finite biochemical reactions within each biochemical pathway of the cellular circuit, forming the basis for intermolecular cross-talks that govern various cellular functions~\cite{kaern2005stochasticity}. The intrinsic fluctuations stemming from random molecular interactions play a pivotal role in regulating cellular organizations~\cite{samoilov2006fluctuations}. These intrinsic fluctuations can be modulated by the system size~\cite{gillespie2007stochastic} (denoted as $V$), with intrinsic fluctuation scaling approximately as $\sim\frac{1}{\sqrt{V}}$~\cite{gillespie2000chemical}. Biological cells can undergo significant changes in size, ranging from a $10$--$50\%$ expansion or shrinkage from their original size. Examples include hippocampal neurons $\sim 10$--$45\%$~\cite{ye2007growing}, erythrocytes (bone marrow cells) $\sim 14$--$20\%$~\cite{canham1968distribution}, and somatic cells $\sim 20$--$40\%$~\cite{shen2021tcf21+}. Intrinsic fluctuations play a crucial role in cellular decision-making and fate determination~\cite{balazsi2011cellular,li2014linear}. Also, the collective behavior of dynamic structures, encompassing steady states, oscillations, chaos, etc., constitutes the foundation of emergent molecular phenotype~\cite{han2008understanding}, shaping the overall phenotypic nature of an organism~\cite{ackermann2015functional}. Apart from simple periodic oscillations, biological systems often exhibit complex oscillations like bursting, birhythmicity, multi-periodicity, quasi-periodicity, and chaos. These phenomena may correspond to complex molecular cross-talks and distributions~\cite{del2002periodicity}. However, the relation between such complex dynamics and molecular cross-talks remains not yet fully understood. In this regard, the impact of intrinsic fluctuations on the complexity of dynamic states poses a fundamental question crucial for gaining deeper insights into the complex mechanisms orchestrated by biological systems. We aim to address this fundamental question through a numerical study of complex oscillations and chaos in a non-linear model of intracellular calcium ion (Ca$^{2+}$) oscillation based on the Ca$^{2+}$-induced Ca$^{2+}$ release (CICR) mechanism proposed by Houart \textit{et al}~\cite{houart1999bursting}. 

Calcium ion (Ca$^{2+}$) functions as a vital messenger within biological cells, undergoing intracellular Ca$^{2+}$ oscillations~\cite{matsu2006cytosolic} in various cell types, including pancreatic cells~\cite{perc2009prevalence,tamarina2005inositol}, hepatocytes~\cite{wu2005phase}, muscle cells~\cite{collier2000calcium, meng2007calcium}, and neurons~\cite{verkhratsky1996calcium}. These oscillations play crucial roles not only in signal transduction inside the cell~\cite{berridge1994spatial,thurley2012fundamental} but also in various physiological processes, including gene expression~\cite{dolmetsch1998calcium}, cell proliferation~\cite{humeau2018calcium}, and neuronal differentiation~\cite{pinto2016studying}. The system of intracellular Ca$^{2+}$ oscillations is a non-equilibrium system \cite{xu2012potential}. Mathematical models have been developed to explain the experimentally observed complex patterns of intracellular Ca$^{2+}$ oscillations. In particular, Houart \textit{et al.}~\cite{houart1999bursting} has developed a model based on the non-linear feedback process of the Ca$^{2+}$-induced Ca$^{2+}$ release (CICR) mechanism~\cite{puebla2005controlling}, prevalent in various cell types, including hepatocytes~\cite{borghans1997complex} and cardiac~\cite{Terrar2020} cells. In the CICR mechanism, the release of  Ca$^{2+}$ from intracellular stores into the cytosol is activated by inositol trisphosphate (InsP$_3$) and cytosolic  Ca$^{2+}$ itself. This autocatalytic process produces a variety of complex dynamical behaviors in the temporal patterns of Ca$^{2+}$ oscillation. Intrinsic stochasticity in intracellular Ca$^{2+}$ oscillations arises due to finite cell size and a small number of reactants~\cite{perc2007periodic}. The interplay of external noise strength and oscillatory dynamics has been studied in the adaption of the biological system of \textit{Physarum polycephalum}~\cite{folz2021interplay}. However, to our knowledge, a systematic analysis of the interplay between intrinsic fluctuation and the ``complexity'' of dynamical behavior such as complex oscillations, including bursting, multi-periodicity, quasi-periodicity, and chaos in the intracellular Ca$^{2+}$ oscillations, has not been addressed earlier. Complexity is a multifaceted notion~\cite{friedrich2011approaching, roli2019complexity}, with various complexity measures proposed across different disciplines, including algorithmic complexity, effective complexity, statistical complexity, fractal dimension, entropy, degree of organization~\cite{lloyd2001measures}, among others. For a deterministic dynamical system described by coupled, non-linear ordinary differential equations and undergoing the bifurcation phenomenon, complexity is generally characterized using Lyapunov exponents~\cite{strogartz1994nonlinear}. In information theory, the degree of disorder is quantified using entropy~\cite{grassberger1991information}. Connecting non-linear dynamics and information theory, Bandt and Pompe have proposed permutation entropy $H$ based on Shannon entropy~\cite{shannon1948mathematical} as a measure of complexity in non-linear time series~\cite{bandt2002permutation}. L$\mathrm{\acute{o}}$pez-Ruiz \textit{et al.}~\cite{lopez1995statistical} have defined a statistical measure of complexity, denoted by $C$, relating order and information. A diagram of $C$ versus $H$, where $H$ is regarded as an arrow of time, is known as the complexity-entropy (CH) causality plane~\cite{rosso2007distinguishing}. In this study, we use permutation entropy $H$ and statistical complexity $C$ as quantitative measures of complexity to analyze the complexities of various dynamical behaviors of intracellular Ca$^{2+}$ and their interplay with intrinsic fluctuations. 

We simulate the dynamics of the intracellular Ca$^{2+}$ oscillation model proposed by Houart \textit{et al.}~\cite{houart1999bursting}, employing a stochastic approach based on the chemical Langevin equation (CLE)~\cite{gillespie2000chemical} to investigate the behavior of the stochastic Ca$^{2+}$ dynamics driven by intrinsic fluctuations. To delve into the intricate relationship between intrinsic fluctuations and the complexity of intracellular Ca$^{2+}$ dynamics, we use permutation entropy~\cite{bandt2002permutation} and the statistical complexity measure~\cite{lopez1995statistical}. We show that intrinsic fluctuations play a significant role in modulating the patterns of Ca$^{2+}$ oscillations within a cell, even to the extent of disrupting the coherence of these oscillations. Applying permutation entropy finds that at large intrinsic fluctuations, states become noisier, and the distinction between the periodic and chaotic states diminishes. Permutation entropy analysis elucidates that the chaotic state is more sensitive to intrinsic fluctuation than the other periodic states. Additionally, we identify distinct periodic and chaotic states of the cytosolic Ca$^{2+}$ at different planar locations on the CH causality plane as intrinsic fluctuation varies, indicating varying complexity and information content. We also observe peaks in the statistical complexities of the different states within an intermediate range of values of intrinsic fluctuation (or system size). Within this regime, we uncover scale-free or fractal patterns in the temporal oscillatory structures of cytosolic Ca$^{2+}$. Our results demonstrate the intricate interplay of intrinsic fluctuation and complexity in the periodic and chaotic states of intracellular Ca$^{2+}$ dynamics, signifying the regulatory role of intrinsic fluctuations within a cell.
  
The structure of the paper unfolds as follows. In Sec.~\ref{sec:two}, we describe the model of intracellular Ca$^{2+}$ oscillations adopted for this study. We outline the methodologies employed, namely, the chemical Langevin equation (CLE) in Sec.~\ref{sec:three_a}, the permutation entropy (PE) in Sec.~\ref{sec:three_b}, and the complexity-entropy (CH) causality plane in Sec.~\ref{sec:three_c}. Sec.~\ref{sec:four} presents and discusses our main results. We then summarize our results with conclusions in Sec.~\ref{sec:five}.

\begin{figure}
\centering
\includegraphics[scale=0.25]{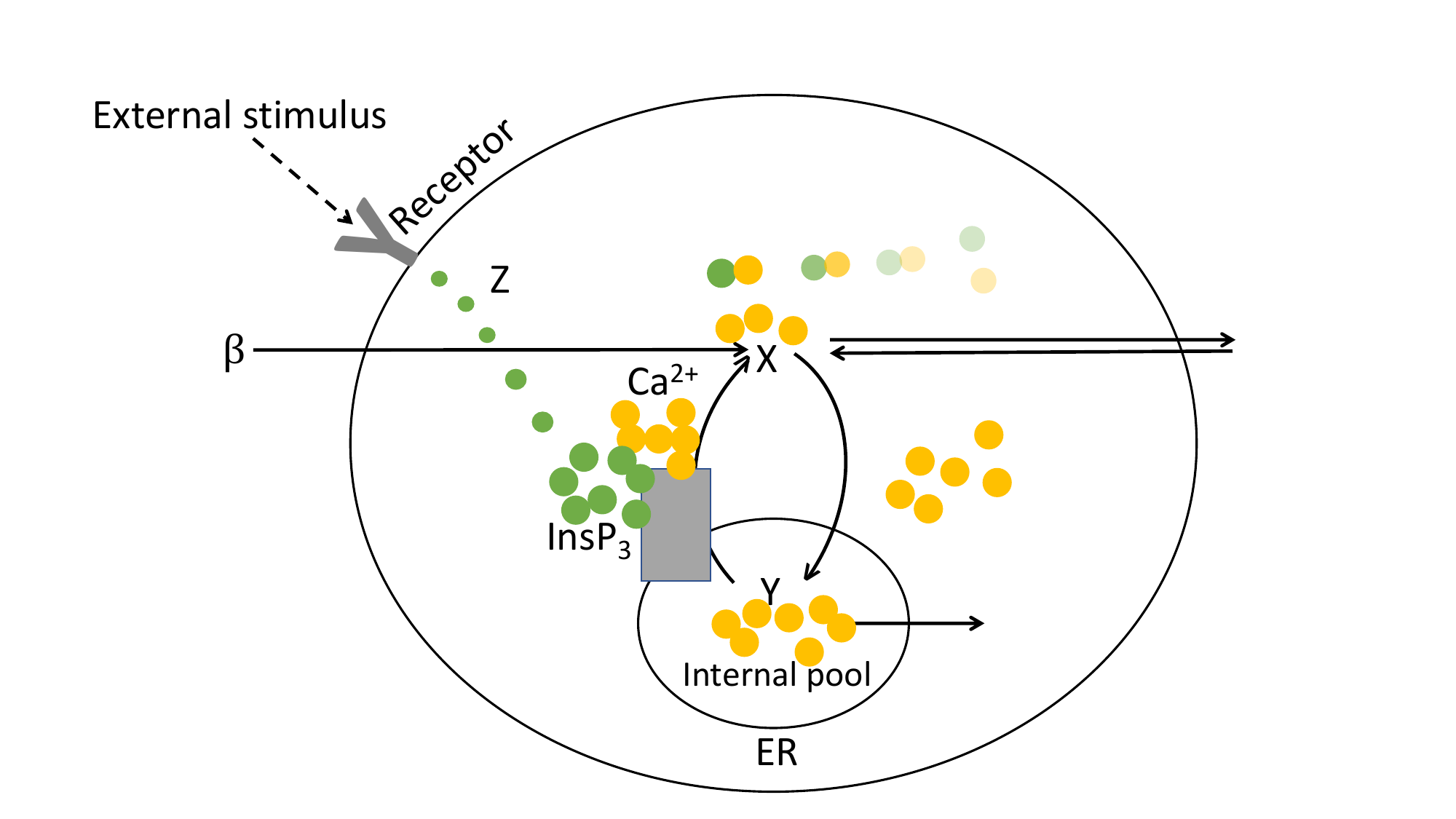}
\caption{Schematic representation of intracellular calcium (Ca$^{2+}$) oscillations based on the interplay between the mechanism of Ca$^{2+}$-induced Ca$^{2+}$-release (CICR) and Ca$^{2+}$-activated 1,4,5-trisphosphate (InsP$_3$) degradation (figure adapted from Ref.~\cite{borghans1997complex}): When an agonist attaches to the plasma membrane receptor, it triggers the synthesis (indicated by dashed arrow) of InsP$_3$ (denoted by $Z$),  an intracellular second messenger. InsP$_3$ binds to receptors on the endoplasmic reticulum (ER) membrane, initiating Ca$^{2+}$-release from the internal pool (denoted by $Y$) into the cytosol through the InsP$_3$ receptor/Ca$^{2+}$ channel (IP$_3$R channel).  This cytosolic Ca$^{2+}$ (denoted by $X$) further activates its own release through the IP$_3$R channel, a phenomenon known as CICR, indicating an autocatalytic process that generates Ca$^{2+}$ oscillations. Stimulation of InsP$_3$ 3-kinase activity is achieved through a Ca$^{2+}$/calmodulin complex.  Dotted arrows depict Ca$^{2+}$-activated InsP$_3$ degradation.  In the figure, $\beta$ represents the degree of cell stimulation by an agonist.  Bold arrows encompass Ca$^{2+}$pumping into the internal pool,  Ca$^{2+}$ released from the pool, Ca$^{2+}$ exchange with the extracellular medium, and stimulus-induced Ca$^{2+}$ influx. Refer to Sec.~\ref{sec:two} for detailed descriptions of the model describing the intracellular Ca$^{2+}$ oscillations. 
}
\label{fig:model}
\end{figure}
\section{Model: Intracellular Calcium ion Oscillations }
\label{sec:two}
Consider a cell of system size $V$ (see Fig.~\ref{fig:model}). The variables $X, Y$, and $Z$ represent the populations of free  Ca$^{2+}$ in the cytosol (cytosolic Ca$^{2+}$), Ca$^{2+}$ stored in the internal pool, and inositol 1,4,5-trisphosphate (InsP$_3$), respectively. Suppose $x=X/V, \ y=Y/V,$ and $z=Z/V$ represent the concentrations of cytosolic Ca$^{2+}$, stored Ca$^{2+}$, and InsP$_3$, respectively. The time-evolution of these concentrations is governed by the following set of coupled, non-linear ordinary differential equations (ODEs)~\cite{houart1999bursting}: 
\begin{align}
\begin{aligned}
\frac{dx}{dt} &=V_0+V_1\beta-V_2+V_3+k_fy-kx, \\
\frac{dy}{dt} &=V_2-V_3-k_fy,   \\
\frac{dz}{dt} &=\beta V_4-V_5-\epsilon z, 
\end{aligned}
\label{eq:model}
\end{align}
where 
\begin{align}
 V_2&=V_{M2} \frac{x^2}{k_2^2+x^2},  \nonumber \\
 V_3&=V_{M3} \frac{x^m}{k_x^m+x^m} \frac{y^2}{k_y^2+y^2} \frac{z^4}{k_z^4+z^4},  \nonumber \\
 V_5&=V_{M5} \frac{z^p}{k_5^p+z^p} \frac{x^n}{k_d^n+x^n}. \nonumber
\end{align}
Intracellular calcium (Ca$^{2+}$) oscillations arise from the interplay between Ca$^{2+}$-induced Ca$^{2+}$-release (CICR) and Ca$^{2+}$-activated 1,4,5-trisphosphate (InsP$_3$) degradation mechanisms (see the caption of Fig.~\ref{fig:model} for detailed description). The coupling of a negative feedback loop (degradation) with a positive CICR cycle gives rise to these oscillations~\cite{borghans1997complex}. In the ODEs~\eqref{eq:model}, the evolution equations  $\frac{dx}{dt}$ and $\frac{dy}{dt}$ of the concentrations $x$ and $y$ encapsulate factors influencing the rate of change in the concentration of cytosolic Ca$^{2+}$ and Ca$^{2+}$ stored, respectively. Here is a breakdown of the terms involved: $V_0$ denotes the constant Ca$^{2+}$ supply from the extracellular medium. The parameter $\beta$ represents the degree of cell stimulation by an agonist (e.g., hormone or neurotransmitter). $V_1$ represents the maximum rate of stimulus-activated Ca$^{2+}$ entry from the extracellular medium. The rate $V_2 \ (V_3)$ corresponds to Ca$^{2+}$ pumping from the cytosol into the internal pool (release of Ca$^{2+}$ from the internal pool to the cytosol). $V_{M2}$ and $V_{M3}$ denote their maximum values. Parameters $k_2, \  k_y, \ k_x$, and $k_z$ represent the threshold values for pumping, release, and activation of release by Ca$^{2+}$ and InsP$_3$, respectively. $V_2$ is solely a function of the cytosolic Ca$^{2+}$ concentration ($x$), whereas $V_3$ depends on all the three concentrations $x,~y$, and $z$. The rate constant $k_f$ measures the passive, linear leak of $y$ into $x$, and $k$ signifies the linear transport of cytosolic Ca$^{2+}$ into the extracellular medium. Regarding $\frac{dz}{dt}$ in the ODEs~\eqref{eq:model}, it encompasses factors affecting the rate of change of InsP$_3$. $V_4$ denotes the maximum rate of stimulus-induced InsP$_3$ synthesis, and $V_5$ represents the phosphorylation rate of InsP$_3$ by the 3-kinase, an InsP$_3$ metabolising enzyme. The decrease of InsP$_3$ is driven by its hydrolysis by calcium-dependent 3-kinase. $k_5$ denotes the half-saturation constant. Stimulation of InsP$_3$ 3-kinase activity (through a Ca$^{2+}$/calmodulin complex) is represented by a Hill-form term with $k_d$ as the threshold level of Ca$^{2+}$. The term $-\epsilon Z$ accounts for the metabolism of InsP$_3$ by 5-phosphatase, independent of Ca$^{2+}$. Additionally, cooperative processes in Ca$^{2+}$ release from internal stores into the cytosol and phosphorylation of InsP$_3$ by 3-kinase are reflected in $V_3$ and $V_5$, incorporating Hill-coefficients $m$, $n$ and $p$. A value of $p>1$ indicates the presence of cooperativity in 3-kinase kinetics, while $p=1$ indicates its absence~\cite{houart1999bursting}. By adjusting the values of the rate constants and other parameters involved in the model~\eqref{eq:model}, concentrations $x$, $y$, and $z$ exhibit various kinds of dynamics, including steady-state, simple periodic oscillations, complex oscillations such as bursting, period doubling sequences, quasi-periodicity, and chaos~\cite{houart1999bursting}.
\section{Methods}
\subsection{Stochastic Modeling with Chemical Langevin Equation (CLE)}
\label{sec:three_a}
Consider a well-stirred chemically reacting system of volume $V$ comprising $N$ molecular species $\{S_1, S_2, \dots,S_N\}$, maintained at a constant temperature $T$. The state vector of molecular populations in the system is denoted by $\textbf{X}(t)=[X_1(t), X_2(t), \dots,X_N(t)]^\mathrm{T}$, where T denotes transpose. Suppose the $N$ species interact through a set of $M$ chemical reactions $\{R_1,R_2,\dots,R_M\}$, where each $R_j$ is represented as: $\ce{$a_{1j}X_1$ + $a_{2j}X_2$ + $\dots$ + $a_{Nj}X_N$ ->[$k_j$] $b_{1j}X_1$ + $b_{2j}X_2$ + $\dots$ +$b_{Nj}X_N$}$. Here, $k_j\ ;\ j=1,2,\dots,M$ denotes the classical rate constant of the $j^{\mathrm{th}}$ reaction, while $\{a_1,a_2,\dots,a_N\}$ and $\{b_1,b_2,\dots b_N\}$ represent the numbers of reactant and product molecules, respectively. The probability of a reaction $R_j$ occurring within $V$ in the next infinitesimal time interval $(t,t+dt)$ is given by $a_j(\textbf{X})dt$, where $a_j(\textbf{X})$ is the propensity function for reaction $R_j$~\cite{gillespie2000chemical}. The function $a_j$ is expressed as $a_j(\textbf{X})=c_jh_j(\textbf{X})$~\cite{gillespie1977exact,gillespie2000chemical}, where $h_j(\textbf{X})$ accounts for possible combinations of molecules for reaction $R_j$. The stochastic rate constant $c_j$ is related to the classical rate constant $k_j$ by $c_j=k_jV^{1-\nu_j}$, with $\nu_j$ being the stoichiometric coefficient in $R_j$. Following Gillespie's formalism~\cite{gillespie2000chemical}, we describe the chemical Langevin equation (CLE) as follows: For an arbitrary time interval $dt>0$, if $\Lambda_j [\textbf{X}(t),dt]$ describes the number of $R_j$ reactions occurring in the subsequent time interval $(t,t+dt)$, then the molecular population $X_i$ at time $t+dt$ is given by
\begin{equation}
   X_i(t+dt)=X_i(t)+\sum_{j=1}^{M} \Lambda_j [\textbf{X}(t),dt] \ \nu_{ji}, \label{eq:cle1}
\end{equation}
where $i=1,2,\dots,N$, and $\nu_{ji}$ is the change in $X_i$ due to $R_j$. Determining $\Lambda_j [\textbf{X}(t),dt]$ is challenging. To obtain an approximation to the function $\Lambda_j$, we impose two important conditions on Eq.~\eqref{eq:cle1}. Firstly, we require the time interval $dt \rightarrow \text{small}$ such that the reaction events occurring during $(t,t+dt)$ do not significantly change the propensity functions, i.e., $\Delta a_j=a_j(\textbf{X}(t'))-a_j(\textbf{X}(t)) \cong 0, \forall \ t'\in (t,t+dt)$. This condition holds when the reactant populations are large ($\gg 1)$, allowing each $\Lambda_j$ to be approximated by a statistically independent Poisson random variable, i.e., $\Lambda_j \rightarrow \mathcal{P}_j(a_j(\textbf{X}),dt)$. Secondly, we require $dt\rightarrow \text{large enough}$ such that the number of $R_j$ reaction events  during $(t,t+dt)$ is far greater than unity, i.e., $a_j(\textbf{X})dt \gg 1$.  This enables each $\mathcal{P}_j(a_j(\textbf{X}),dt)$ to be approximated by a normal random variable such that $\mathcal{P}_j(a_j(\textbf{X}),dt) \rightarrow \mathcal{N}_j(a_j(\textbf{X})dt, a_j(\textbf{X})dt)$. These two conditions are applied simultaneously in the limit of a large molecular population, rendering $dt$ \textit{macroscopically infinitesimal}~\cite{gillespie2000chemical}. 

With $\mu$ and $\sigma$ as the mean and standard deviation, respectively, we use $\mathcal{N}(\mu,\sigma^2)=\mu+\sigma \mathcal{N}(0,1)$~\cite{gillespie1996multivariate} to get   
 \small{
\begin{equation}
    X_i(t+dt)= X_i(t)+\sum_{j=1}^{M} \nu_{ji} \ a_j(\textbf{X})dt+\sum_{j=1}^{M}\nu_{ji}\ [a_j(\textbf{X})dt]^{1/2}\ \mathcal{N}_j(0,1).\label{eq:sfle}
\end{equation}
}
\normalsize
Eq.~\eqref{eq:sfle} represents the standard-form Langevin equation~\cite{gillespie1996multivariate,gillespie2000chemical}. On the right-hand side, the term $\nu_{ji} \ a_j(\textbf{X})dt$ represents a deterministic component, while $\nu_{ji}\ a_j^{1/2}(\textbf{X})\ \mathcal{N}_j(0,1)dt^{1/2}$ denotes a stochastic component. 

Algebraically rearranging Eq.~\eqref{eq:sfle}, we get 
\begin{align}
    &\frac{X_i(t+dt)-X_i(t)}{dt}\nonumber \\
    &=\sum_{j=1}^{M} \nu_{ji} \ a_j(\textbf{X})+\sum_{j=1}^{M}\nu_{ji}\ a_j^{1/2}(\textbf{X})\mathcal{N}_j(0,1)dt^{-1/2},
\end{align}
where $\mathcal{N}(0,1)dt^{-1/2}=\mathcal{N}(0,1/dt)$~\cite{gillespie1996multivariate}.
Taking the limit $dt\rightarrow 0$~\cite{gillespie1996multivariate,gillespie2000chemical}, we arrive at CLE with a general form:
\begin{equation}
    \frac{d X_i(t)}{dt}=\sum_{j=1}^{M} \nu_{ji} \ a_j(\textbf{X})+\sum_{j=1}^{M}\nu_{ji}\ a_j^{1/2}(\textbf{X})\ \xi_j(t), \label{eq:cl}
\end{equation}
where $\xi_j(t)$ are temporally uncorrelated, statistically independent Gaussian white noises with $\xi_j(t)=\displaystyle \lim_{dt\rightarrow 0} \mathcal{N}(0,1/dt)$~\cite{gillespie1996multivariate,gillespie2000chemical}. On the right-hand side, the first term accounts for the deterministic part, whereas the second term represents the stochastic term. The CLE~\eqref{eq:cl} thus accounts for both deterministic and stochastic components in the dynamics of a chemically reacting system.

In Eq.~\eqref{eq:cl}, the drift and diffusion components both follow from the Poisson random variable $\mathcal{P}_j$~\cite{gillespie2000chemical}. Hence, the ratio of the random to the deterministic component is $a_j^{-1/2}(\textbf{X})$. For a Poisson distribution, the standard deviation $\sigma =\sqrt{\mu}$. Assuming the average number of molecules $\mu$ is proportional to the system size $V$, we have $\sigma \propto \sqrt{V}$. Consequently, the magnitude of intrinsic fluctuations in molecular populations, given by $(1/\sigma)$, scales as $(1/\sqrt{V})$. Thus, for stochastic chemical reactions described by Poisson random variable $\mathcal{P}_j$, intrinsic fluctuations in molecular populations scale as the inverse square root of the system size $V^{-1/2}$~\cite{gillespie2000chemical}.   

In the thermodynamic limit, as the number of molecules $N\rightarrow \infty$ and the system size $V \rightarrow \infty$ such that the species concentration $(N/V) \rightarrow \text{constant}$, intrinsic fluctuations vanish. Consequently, Eq.~\eqref{eq:cl} becomes
\begin{equation}
    \frac{d X_i(t)}{dt}=\sum_{j=1}^{M} \nu_{ji} \ a_j(\textbf{X}),
\end{equation}
which represents the macroscopic (deterministic) reaction rate equation of traditional chemical kinetics. Deterministic reaction rate equations are generally used to explain the complex nonlinear dynamics observed in biological systems. However, the dynamics of small biological systems are greatly influenced by fluctuations. Hence, it is important to investigate the effect of these fluctuations on the deterministic nonlinear dynamics. In the case of Ca$^{2+}$ oscillations, experimental observations have revealed that the complex intracellular Ca$^{2+}$ oscillations exhibit stochastic dynamics~\cite{green1993adenine,marrero1994taurolithocholate,borghans1997complex}. Although Houart \textit{et al.}~\cite{houart1999bursting} have successfully explained the complex Ca$^{2+}$ oscillations through deterministic analysis using the nonlinear model~\eqref{eq:model}, it is crucial to consider the inherent stochasticity that may arise from system size variation, as discussed in Introduction. This intrinsic fluctuation, whose strength scales inversely with the square root of the system size, needs to be accounted for to provide a more realistic representation. Hence, we adopt stochastic modeling via the chemical Langevin equation (CLE) to capture the experimentally observed stochastic dynamics and to explore the impact of intrinsic fluctuation on the complex nonlinear dynamics of intracellular Ca$^{2+}$ oscillations.
\subsection{Permutation Entropy (PE)}
\label{sec:three_b}
The permutation entropy (PE) method, developed by Bandt and Pompe~\cite{bandt2002permutation}, serves as a natural complexity measure for time series data. The Bandt-Pompe approach involves symbolizing a time series by converting it into symbolic sequences. It determines a probability distribution known as the ordinal probability distribution for ordinal or permutation patterns in the sequence. The ordinal probability distribution captures the likelihood of observing different ordinal patterns, providing valuable insights into the complexity of the time series. PE is then computed as the Shannon entropy of this ordinal probability distribution. PE is based on measuring the information contained in comparing some $r$ consecutive values (known as the order of permutation or embedding dimension) in a time series. 

We explain the Bandt-Pompe procedure in the following. It begins by dividing any giventime series $\mathcal{X}$ of length $N$, denoted by $\mathcal{X}=\{x_i~;~i=1,2,\dots,N\}$, into overlapping partitions $n=N-(r-1)\tau$, where $\tau$ represents the embedding delay~\cite{pessa2021ordpy}. For each data partition $D_p=(x_p,x_{p+\tau},\dots,x_{p+(r-1)\tau})$ with $p=1,\dots,n$ as the partition index, a permutation $\pi_p=(s_0,s_1,\dots,s_{r-1})$ of $(0,1,\dots,r-1)$ is determined by sorting the elements in ascending order. The permutation of the index numbers is defined by the inequality $ x_{p+s_0}\leq x_{p+s_1}\leq \dots \leq x_{p+s_{r-1}}$~\cite{bandt2002permutation}. The final symbolic sequence is given by $\{\pi_p\}_{p=1,\dots,n}$. The relative frequency of all possible patterns in the symbol sequences is ~\cite{bandt2002permutation}
\begin{equation}
    \rho_j(\pi_j)= \frac{\hbox{
 \# patterns of type}~\pi_j~\hbox{in permutation}~\{ \pi_p\}}{n}.
\end{equation}
Then the permutation entropy $S[P]$ is defined as~\cite{bandt2002permutation}
\begin{equation}
    S[P]=-\sum_{j=1}^{r!} \rho_j(\pi_j) \log \rho_j(\pi_j), \label{eq:pe1}
\end{equation}
where the ordinal probability distribution $P=\{\rho_j(\pi_j)\} \ ; \ j=1,\dots,r!$. The permutation entropy per symbol of order $r$, denoted by $h_r$, is then given by \cite{bandt2002permutation,riedl2013practical}
\begin{equation}
h_r=\frac{S[P]}{r-1}. \label{eq:pe2}
\end{equation} 
Here, the normalization factor ($r-1$) accounts for the fact that ($r-1$) pairwise comparisons are present in the formation of each ordinal pattern.

For example, consider the time series $\mathcal{X}=(6,2,10,8)$ with an embedding dimension of $r=3$. Hence, $r!=3!=6$ possible permutations $\{\pi_p\}$ arise, namely $\pi_1=(0,1,2), \pi_2=(0,2,1), \pi_3=(1,0,2), \pi_4=(1,2,0), \pi_5=(2,0,1)$ and $\pi_6=(2,1,0)$. Suppose we choose consecutive time units, i.e., $\tau=1$. The first data partition $D_1=(6,2,10)$ corresponds to $(x_t,x_{t+1},x_{t+2})$. Sorting the elements in ascending order yields $2<6<10$, indicating $x_{t+1}<x_{t}<x_{t+2}$. Therefore, the ordinal pattern associated with $D_1$ is $\pi_3=(1,0,2)$. Moving to the second data partition $D_2=(2,10,8)$, sorting the elements in ascending order gives $x_{t}<x_{t+2}<x_{t+1}$, implying the ordinal pattern is $\pi_2=(0,2,1)$. 

\subsection{Complexity-Entropy (CH) causality plane}
\label{sec:three_c}
L$\mathrm{\acute{o}}$pez-Ruiz \textit{et al.} ~\cite{lopez1995statistical} have introduced the concept of complexity ($C$) of a system as the product of disequilibrium ($DE$) and entropy ($H$):
\begin{equation}
    C=DE\times H. \label{eq:stat}
\end{equation}
Here, $H$ measures the information the system stores, while $DE$ represents the system's deviation from an equiprobable distribution. In essence, complexity $C$ reflects the interplay between the information stored in a system (quantified by $H$) and the system's departure from equiprobability (quantified by $DE$). We note that entropy measures disorder, while disequilibrium measures order \cite{smaal2021complexity}.  

Building upon this concept, Rosso \textit{et al.} ~\cite{rosso2007distinguishing} have further defined the statistical complexity measure $C$ as
\begin{equation}
    C=DE[P,U]\ H[P], \label{eq:scm}
\end{equation}
where $P$ and $U$ represent the ordinal and uniform probability distributions, respectively. The uniform distribution is given by $U=\{1/r!, \ldots, 1/r!\}$.   

Connecting with the permutation entropy $S[P]$ defined in Eq.~\eqref{eq:pe1}, the normalized permutation entropy $H$ is defined as~\cite{rosso2007distinguishing}
\begin{equation}
    H=\frac{S[P]}{\log r!}. \label{eq:npe}
\end{equation}
The normalization factor $\log r!$ reflects the maximum entropy such that $0 \leq H\leq 1$.

In Eq.~\eqref{eq:scm}, the disequilibrium $DE$ is expressed as $DE[P,U]=D_0D$, where $D=S\bigg[\frac{(P+U)}{2}\bigg]-\frac{S[P]}{2}-\frac{S[U]}{2}$ is the Jensen-Shannon divergence~\cite{martin2006generalized} and the normalization constant $D_0=-\frac{1}{2}\bigg [ \left(\frac{r!+1}{r!}\right)- 2 \log(2r!) +\log r!\bigg ]$ \cite{kowalski2005entropic,zunino2007characterization}.

The complexity-entropy (CH) causality plane, a two-dimensional causality plane, visually represents the values of the statistical complexity measure $C$ and the normalized permutation entropy $H$. This plane was used to distinguish stochastic and chaotic time series~\cite{rosso2007distinguishing}. While the normalized permutation entropy $H$ only measures disorder ($H=0$ for complete order and $H=1$ for complete disorder), the statistical complexity measure $C$ of Eq.~\eqref{eq:scm} quantifies both randomness and the degree of correlational structures~\cite{rosso2007distinguishing}, making it a powerful tool for analyzing complexity in time series. Specifically, $C\simeq 0$ indicates both regular and completely random time series. 
\section{Results and Discussion}
\label{sec:four}
We now present the results and discuss our analyses in the following subsections. 
\subsection{Regulation of the patterns of cytosolic Ca$^{2+}$ dynamics by intrinsic fluctuations}
\begin{figure}
\centering
\includegraphics[scale=0.5]{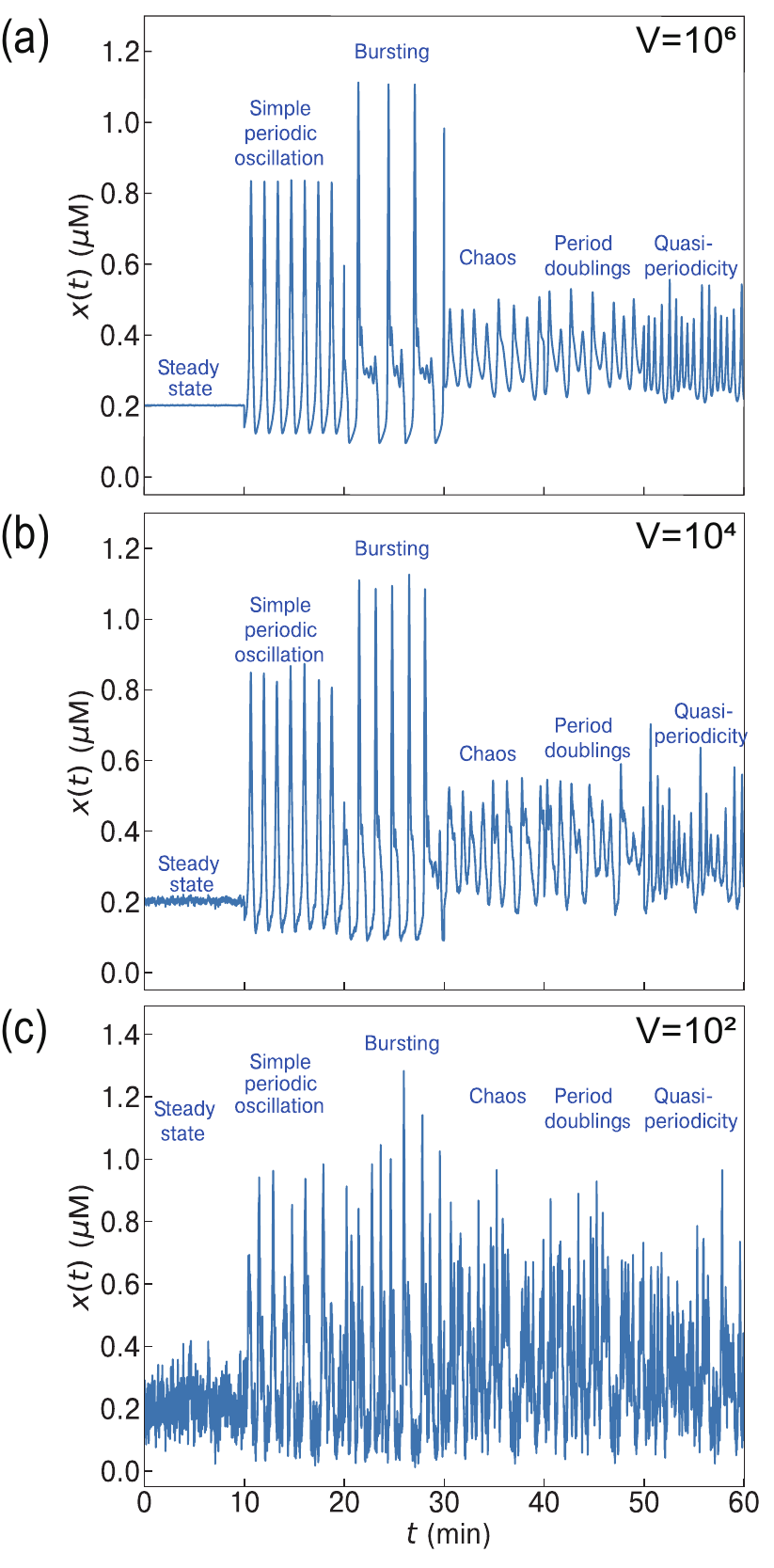} 
\caption{Time evolution of the concentration $x(t)$ of the cytosolic Ca$^{2+}$ by solving the chemical Langevin equation (CLE)~\eqref{eq:cle} for the system size: (a) $V=10^6$, (b) $V=10^4$, and (c) $V=10^2$. Fluctuation-driven dynamics are observed in the distinct dynamic states of the cytosolic Ca$^{2+}$, namely, steady-state, simple periodic oscillation, complex oscillations such as bursting, period doubling sequences, and quasi-periodicity, and chaos. }
\label{fig:oscillations}
\end{figure}
\begin{figure*}
\includegraphics[width=\textwidth]{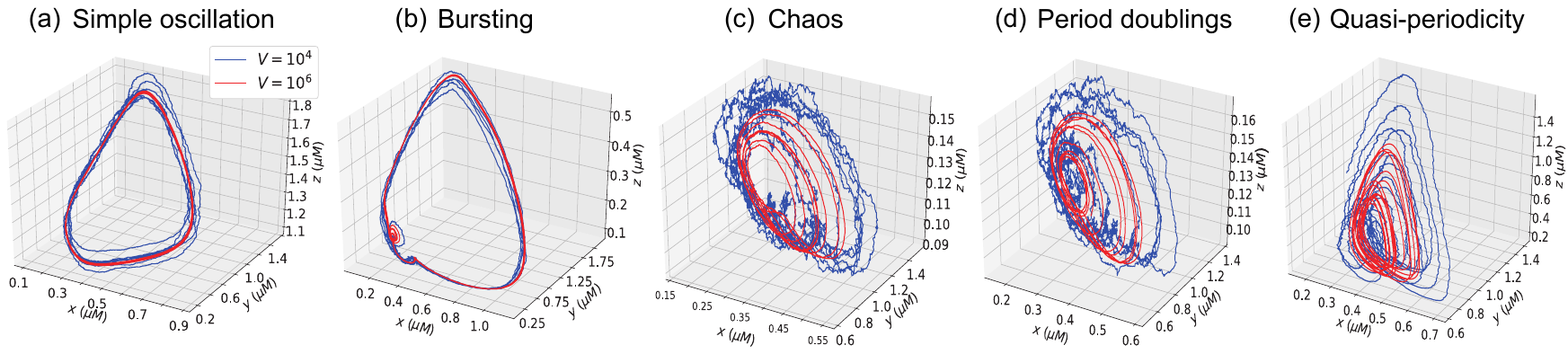}
\vspace{-8cm}
\caption{Phase space plots of the different dynamics of the intracellular Ca$^{2+}$ oscillation model~\eqref{eq:model}, namely, (a) simple periodic oscillation, (b) bursting, (c) chaos, (d) period doubling sequences, and (e) quasi-periodicity for systems of $V=10^4$ (blue curves) and $10^6$ (red curves). For clear visualization of the states of chaos and period doubling sequences, their phase space plots are redrawn for $V=10^{10}$ in Fig.~\ref{fig:append}.}
\label{fig:pp}
\end{figure*}
To capture the experimentally observed stochastic dynamics of intracellular Ca$^{2+}$ oscillations, we adopt a stochastic approach to the non-linear model~\eqref{eq:model} using the chemical Langevin equation~\eqref{eq:cl}. The stochastic approach will allow us to analyze the impact of intrinsic fluctuations on the dynamics of Ca$^{2+}$. Representing $x, y$, and $z$ as the concentrations of cytosolic Ca$^{2+}$, Ca$^{2+}$ stored in the internal pool, and InsP$_3$, respectively, we denote the state vector of concentrations as $s=s(t)=[x(t),~y(t),~z(t)]^{\mathrm{T}}$. T denotes transpose. We then rewrite the coupled, non-linear ODEs~\eqref{eq:model} as
\begin{equation}
    \frac{ds}{dt}=F(x,y,z), \label{eq:ten}
\end{equation}
where 

\begin{equation}
    F(x,y,z)=\begin{bmatrix}
    V_0+V_1\beta-V_2+V_3+k_fy-kx\\
    V_2-V_3-k_fy\\
    \beta V_4-V_5-\epsilon z
    \end{bmatrix}. \nonumber 
\end{equation}
Using stochastic modeling, we now represent the system of ODEs.~\eqref{eq:ten} as a set of reaction channels, illustrating the changes of the populations $X, Y$, and $Z$ (see Table~\ref{table:one} in Appendix~\ref{app:A1}). The state transitions reflect random births and deaths of the molecular species $X, Y$, and $Z$, introducing intrinsic fluctuations in their populations. We determine the propensity functions of each reaction channel using the definition in Sec.~\ref{sec:three_a}. Using Eq.~\eqref{eq:cl}, we determine the chemical Langevin equation (CLE) for the intracellular Ca$^{2+}$ oscillation model~\eqref{eq:ten} as:
\begin{equation}
    \frac{ds}{dt} =F(x,y,z)+\frac{1}{\sqrt{V}} \ G(x,y,z),\label{eq:cle} 
    \end{equation}\\
where $V$ denotes the system size. $G(x,y,z)$ is:
   \small
\begin{align}
    &G(x,y,z)\nonumber \\
    &=\begin{bmatrix}
    \sqrt{V_0}\xi_1+\sqrt{V_1 \beta}\xi_2-\sqrt{V_2}\xi_3+\sqrt{V_3}\xi_4 +\sqrt{k_fy}\xi_5-\sqrt{kx}\xi_6\\
    \sqrt{V_2}\xi_7-\sqrt{V_3}\xi_8-\sqrt{k_fy}\xi_9\\
    \sqrt{V_4 \beta}\xi_{10}-\sqrt{V_5}\xi_{11}-\sqrt{\epsilon z}\xi_{12}
    \end{bmatrix}. \label{eq:noise}
    \end{align}
\normalsize
Here, $\xi_j$ ($j=1,2,\dots,12$) represents a statistically independent Gaussian white noise with the properties of $\langle \xi_j(t) \rangle=0$ and $\langle \xi_j(t) \xi_{j'}(t')\rangle=\delta_{jj'} \delta(t-t').$ 
The second term in Eq.~\eqref{eq:cle} represents the stochastic component, with 
$\frac{1}{\sqrt{V}}$ accounting for the effects of system size or intrinsic fluctuation on the Ca$^{2+}$ oscillation dynamics~\cite{zhu2007mesoscopic}. The parameter $V$ controls the strength of intrinsic fluctuations. In the thermodynamic limit ($V\rightarrow \infty$), intrinsic fluctuations vanish, and the stochastic differential equation~\eqref{eq:cle} becomes the deterministic equation~\eqref{eq:ten}. 

We solve the CLE~\eqref{eq:cle} using the Euler method with a time step of $0.001$ minutes and a total of $10^4$ observations. Time is measured in minutes (min), the concentration of cytosolic Ca$^{2+}$, $x(t)$ in $\mu$M and $V$ in the unit of $\mu$m$^3$. For reference, we provide some physiological ranges of Ca$^{2+}$ concentrations in biological cells. While the extracellular Ca$^{2+}$ concentration is $\sim$1--3 mM, the intracellular Ca$^{2+}$ concentration is relatively low ($\sim$ 0.1--0.2$~\mu$M)~\cite{BLANCO2017547,WEAVER2020321}. Depending on the cell type, cytosolic Ca$^{2+}$ concentration can vary, for instance, $\sim$ 0.1--0.2 $\mu$M in hepatocytes~\cite{murphy255rich}, $\sim 0.26 ~\mu$M in rat myocytes~\cite{williamson1983cytosolic}, and $\sim$ 0.25--0.3~$\mu$M for cardiac muscle~\cite{williamson1983cytosolic}. Depending on the parameter values (see Table~\ref{table:two} of Appendix~\ref{app:Ae}), cytosolic Ca$^{2+}$ concentration $x(t)$ shows various dynamical states, including the steady state, simple periodic oscillation, and complex oscillations such as bursting, period doubling sequences, quasi-periodicity, and chaos. We refer the readers to Ref.~\cite{houart1999bursting} for detailed characterization of these dynamic states using bifurcation diagrams, Lyapunov exponents, first-return maps, and power spectra. 

 Fig.~\ref{fig:oscillations} presents the time evolution of the cytosolic Ca$^{2+}$ concentration, $x(t)$ for three different system sizes: $V=10^6$, $10^4$, and $10^2$. Correspondingly, the strength of intrinsic fluctuation, measured by $~\frac{1}{\sqrt{V}}$, varies approximately as $\sim 0.001$, $0.01$ and $0.1$ for $V=10^6$, $10^4$, and $10^2$, respectively. The concentration $x(t)$ in each panel of Fig.~\ref{fig:oscillations} is generated by merging six distinct time series obtained separately by solving the CLE~\eqref{eq:cle} under six different sets of parameter settings corresponding to each dynamic state (see Table~\ref{table:two} of Appendix~\ref{app:Ae}). For a more extended simulation, see Fig.~\ref{fig:appendix_figure0} in Appendix~\ref{app:A1}. Intrinsic fluctuation regulates the dynamical behavior of cytosolic Ca$^{2+}$ concentration depending on the system size. 

In Fig.~\ref{fig:oscillations}(a), the cytosolic Ca$^{2+}$ concentration $x(t)$ exhibits small fluctuations at system size $V=10^6$. When $V$ becomes larger, the CLE~\eqref{eq:cle} will approach $\frac{ds}{dt}\approx F(x,y,z)$, showing a transition towards the deterministic limit. Such a transition from stochastic to deterministic dynamics at large system size is often called the thermodynamic limit~\cite{gillespie2002chemical}.

In Fig.~\ref{fig:oscillations}(b), discernible fluctuations appear in the dynamics of $x(t)$ at $V=10^4$. As $V$ decreases, the growing intrinsic fluctuation significantly impacts the patterns of periodic and chaotic states of $x(t)$. The state of periodic doubling sequences appears to lose its double periodicity and exhibits behavior akin to chaos at a large value of intrinsic fluctuation. Previously, intrinsic noise was found to impart chaoticity to periodic limit cycles in the chemical Lorentz system~\cite{thounaojam2022stochastic}.   

In Fig.~\ref{fig:oscillations}(c), intrinsic fluctuations have a pronounced effect at the small system size $V=10^2$, obliterating the discernible dynamic patterns in $x(t)$. The large intrinsic fluctuations have obscured the complex oscillations such as bursting, period doubling sequences, and quasi-periodicity that were previously differentiable. Chaos becomes no longer distinguishable from noise. 

In Fig.~\ref{fig:pp}, we plot the phase diagrams for various oscillatory dynamics introduced in Fig.~\ref{fig:oscillations} at $V=10^4$ (blue) and $10^6$ (red). The phase diagrams reveal distinct cyclic patterns inherent to each dynamic state when the system is large ($V=10^6$). We observe a limit cycle for simple periodic oscillation, a limit cycle with multiple small loops for bursting, a strange attractor for chaos, a double loop for period doubling sequences, and a torus for quasi-periodicity. Refer to Fig.~\ref{fig:append} to see chaos and period-doubling sequences clearly. When intrinsic fluctuations become larger for $V=10^4$ (blue), the fine structure in the phase diagram becomes blurred while the large-scale structure is maintained. If the system size is further decreased to $V=10^2$, noise dominates the deterministic dynamics, and the phase diagram becomes completely noisy (not shown).      
\subsection{Interplay of the complexity of cytosolic Ca$^{2+}$ with intrinsic fluctuation using permutation entropy}
\label{sec:pe}
\begin{figure}
\includegraphics[scale=0.5]{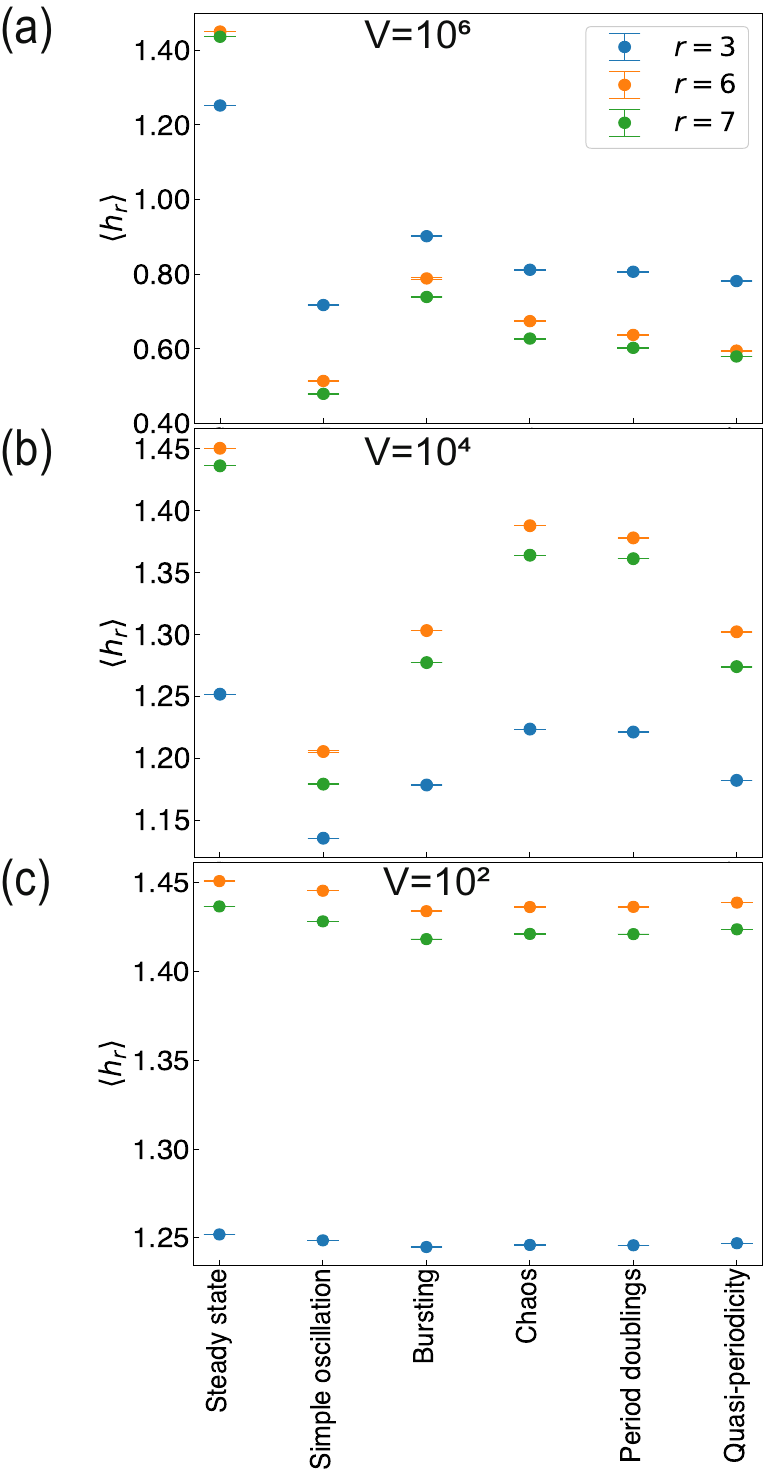}
\vspace{-0.5cm}
\caption{The mean permutation entropy $\langle h_r \rangle$ per symbol of order $r$ for various dynamic states of cytosolic Ca$^{2+}$ at (a) $V=10^6$, (b) $V=10^4$, and (c) $V=10^2$. For each data point on the plot, the ensemble-average was evaluated with 600 time series [$x(t)$] of length $10^4$ after removing the initial 500 data points to avoid transient effects. The error bars represent standard errors. It is observed that the values in $\langle h_r \rangle$ vary across different system sizes and embedding dimensions, which reveals the complex interplay between the cytosolic Ca$^{2+}$ concentration and the system size or intrinsic fluctuations.
}
\label{fig:peavg}
\end{figure}
For calculating the permutation entropy with the embedding dimension $r$, denoted as $h_r$ (see the definition in Eq.~\eqref{eq:pe2}), we consider three values of $r=3$, $6$, and $7$ in line with practical recommendations~\cite{bandt2002permutation}. The numerical code for calculating $h_r$ is implemented in Fortran90. Here the permutation entropy is calculated with the $x$-component of the state vector $s(t)$ in Eq.~\eqref{eq:ten}, i.e., the time series of the cytosolic Ca$^{2+}$ concentration.

In Fig.~\ref{fig:peavg}, we calculate the ensemble-averaged permutation entropy $\langle h_r \rangle$ for various dynamic states of cytosolic Ca$^{2+}$ at (a) $V=10^6$, (b) $V=10^4$, and (c) $V=10^2$ (see also the time evolution of $h_r$ from a single time series in Fig.~\ref{fig:pe}). Each data point was evaluated with 600 distinct $x(t)$ of length $10^4$. The error bars represent standard errors. The permutation entropy captures the distinctive features of different dynamic states in cytosolic Ca$^{2+}$, with higher values indicating higher disorder or complexity. Let us start with the case for $V=10^6$. First, we observe that the steady state has the highest value in $\langle h_r \rangle$ among the six states. This can be explained by the fact that the steady state is dominated by random fluctuations without any ordered patterns (see also Fig.~\ref{fig:appendix_figure0}(a)). Following the steady state, the bursting state exhibits moderate disorder resulting from fine bursting structures, including a number of spikes with reduced amplitudes around the plateau (see dynamical structure of bursting in ~Fig.~\ref{fig:oscillations}(a)). Next to the steady state and bursting, the chaos state exhibits the third highest $\langle h_r \rangle$ value. This suggests that a higher degree of order or regularity is inherent in the chaotic dynamics when observed through permutation entropy. Simple periodic oscillation has the lowest entropy being the simplest dynamical structure among the oscillatory structures considered (Fig.~\ref{fig:pp}), implying the least complexity.

When the system size is reduced to $V=10^4$ [Fig.~\ref{fig:peavg}(b)],
compared to $V=10^6$, the values of $\langle h_r \rangle$ across all periodic and chaotic states become elevated. The Ca$^{2+}$ dynamics attain increased disorder driven by a larger intrinsic fluctuation $(\sim 0.01)$. Particularly noteworthy is the amplified impact of intrinsic fluctuations when interacting with the states of chaotic and period-doubling sequences. This behavior is understood such that chaos generally arises and vanishes via a period-doubling cascade, obeying a sequence of bifurcations known as Feigenbaum sequence~\cite{houart1999bursting, pomeau1984order} in non-linear dynamical systems. The observed higher impact of internal fluctuations on chaos aligns with the previous report by Wu and Kapral~\cite{wu1993internal} in the Willamowski-R$\mathrm{\ddot{o}}$ssler model for deterministic chemical chaos~\cite{willamowski1980irregular}.

In Fig.~\ref{fig:peavg}(c), we plot $\langle h_r \rangle$ of $V=10^2$. In such a small system with strong intrinsic fluctuations, the $\langle h_r \rangle$ values of the different dynamic states of the cytosolic Ca$^{2+}$ concentration are almost the same, signifying a transition from order to disorder state. As also previously seen in Fig.~\ref{fig:oscillations}(c), the large internal fluctuations destroy the distinctive features in the periodic and chaotic patterns of the cytosolic Ca$^{2+}$ concentration. 

We note that in Figs.~\ref{fig:peavg} (and A.3) the relative magnitude of permutation entropy among the six dynamic states of cytosolic Ca$^{2+}$ remains consistent for varying embedding dimension $r$. Increasing $r$ results in the effect of pronounced differences in $\langle h_r \rangle$ among the dynamic states. A higher embedding dimension provides a better resolution of the differences in the complexities of the different states. This is because higher $r$ captures more temporal patterns in the time series $x(t)$. Another notable observation is that for a given embedding dimension $r$, the permutation entropy tends to increase as $V$ decreases. This is due to the fact that the intrinsic fluctuation, quantified by $\frac{1}{\sqrt{V}}$, increases the disorder in the cytosolic Ca$^{2+}$ concentration $x(t)$ as the system size becomes smaller. Importantly, for all $r$ values investigated, we find that $\Delta h_r^{\mathrm{II}}<\Delta h_r^{\mathrm{I}}$, where $\Delta h_r^\mathrm{I}=h_r(V=10^4)-h_r(V=10^6)$ and $\Delta h_r^{\mathrm{II}}=h_r(V=10^2)-h_r(V=10^4)$. This behavior suggests that higher intrinsic fluctuations (or smaller $V$) reduce the relative difference in the permutation entropy. 
\begin{figure}
\includegraphics[scale=0.3]{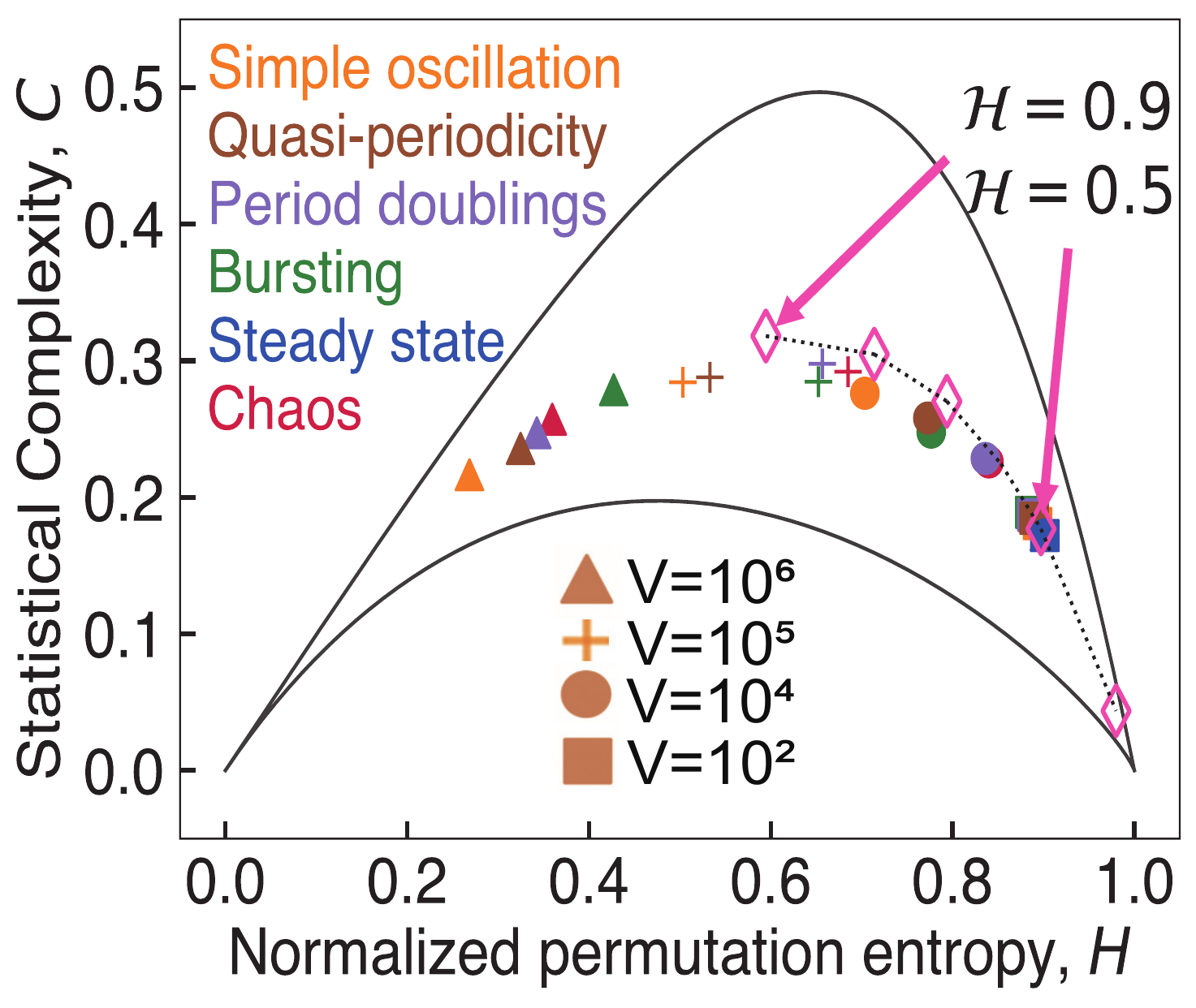}
\caption{Dynamic states of the cytosolic Ca$^{2+}$ in the complexity-entropy (CH) causality plane. 
Here, the statistical complexity measure $C$ and the normalized permutation entropy $H$ were calculated using the open-source Python module \texttt{ordpy}~\cite{pessa2021ordpy}. Six distinct dynamic states are marked on this plane for $V=10^2$ (squares), $10^4$ (circles), $10^5$ (plus signs), and $10^6$ (triangles). The solid lines represent the theoretical curves of the maximum and minimum values of statistical complexity with the embedding dimension $r=6$. The unfilled diamonds mark the $(H,C)$ values for fBm processes at several Hurst exponent $\mathcal{H}$. The
black dotted line is the guideline. The plot reveals a non-monotonic relationship between the statistical complexity and the permutation entropy of cytosolic Ca$^{2+}$ dynamics.} 
\label{fig:plane}
\end{figure}
\subsection{Interplay of the complexity of cytosolic Ca$^{2+}$ with intrinsic fluctuation using complexity-entropy ($CH$) causality plane}
\label{sec:ceplane}
\begin{figure*}
\includegraphics[width=\textwidth]{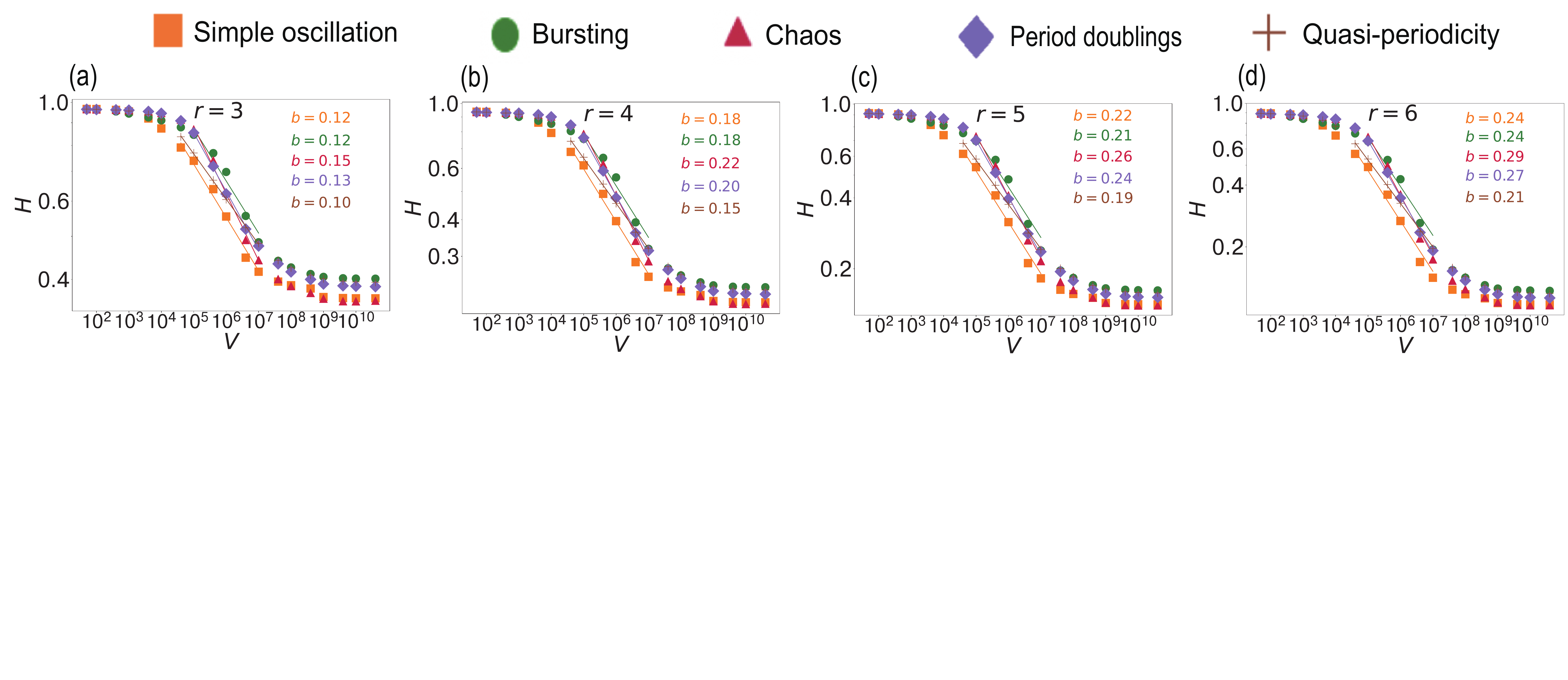}
\vspace{-4cm}
\caption{(a--d) Log-log plots of the normalized permutation entropy $H$ as a function of $V$ at $r=3,~4,~5$, and $6$. The markers show the data for the five distinct cytosolic Ca$^{2+}$ states.  The solid lines are the fits to the data in $V\in [10^5,~10^7]$ using a power-law function $H(V)=aV^{-b}$. The estimated values of the power-law exponent $b$ are listed in each panel.}
\label{fig:ecplane1}
\end{figure*}
\begin{figure}
\includegraphics[scale=0.24]{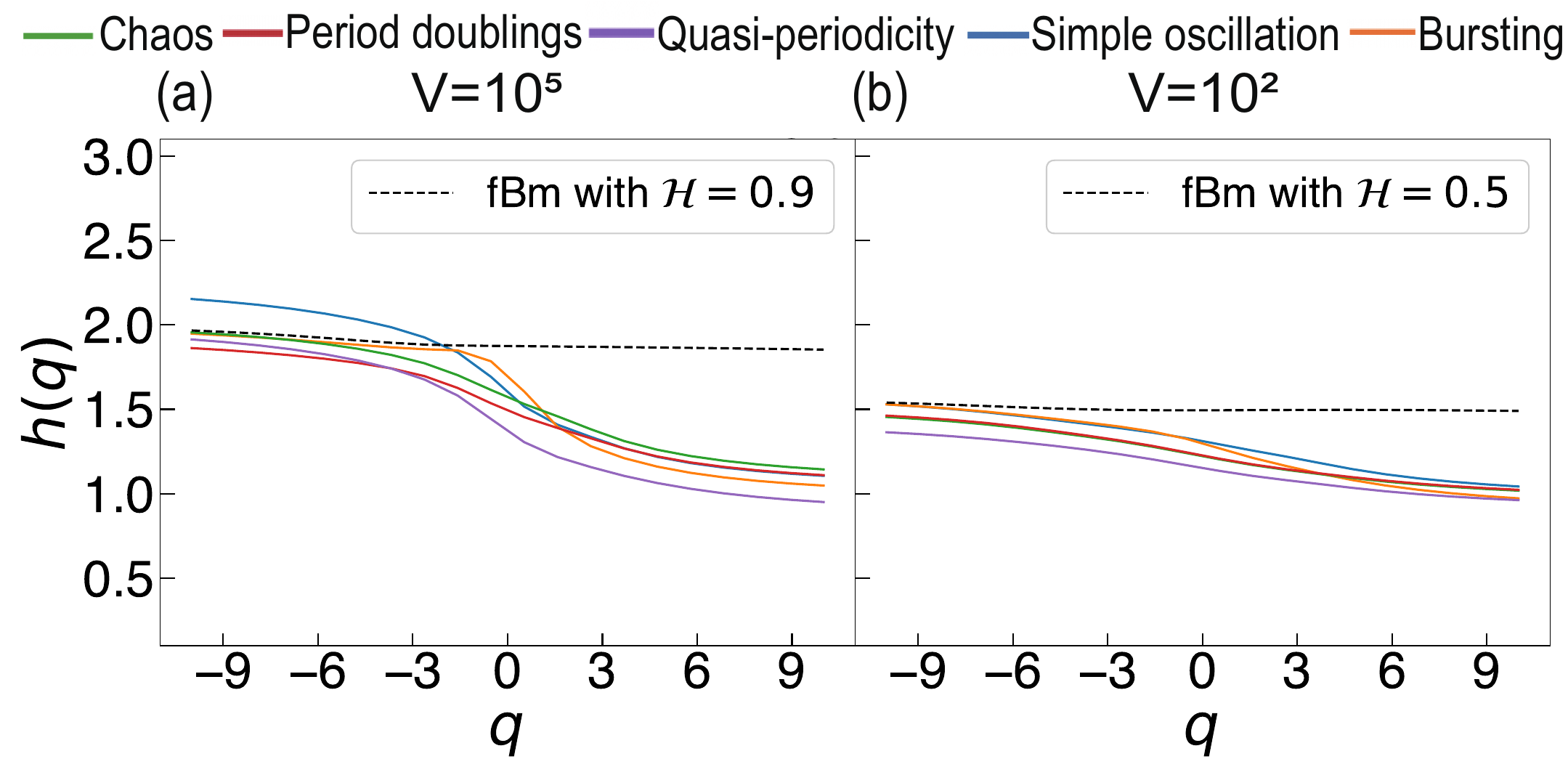}
\caption{Comparison of the generalized Hurst exponent $h(q)$ of the dynamical states (solid lines) of Ca$^{2+}$ with that (dashed lines) of fractional Brownian motion (fBm): the five states of Ca$^{2+}$ (a) for $V=10^5$ with fBm of $\mathcal{H}=0.9$, and (b) for $V=10^2$ with fBm of $\mathcal{H}=0.5$. The variation of $h$ over $q$ indicates that the time series of Ca$^{2+}$ concentrations feature scale-dependent fractality (multifractality). In contrast, fBm is a  monofractal process, yielding a constant $h$ value over $q$. }
    \label{fig:fbm}
\end{figure}
\begin{figure*}
\includegraphics[width=\textwidth]{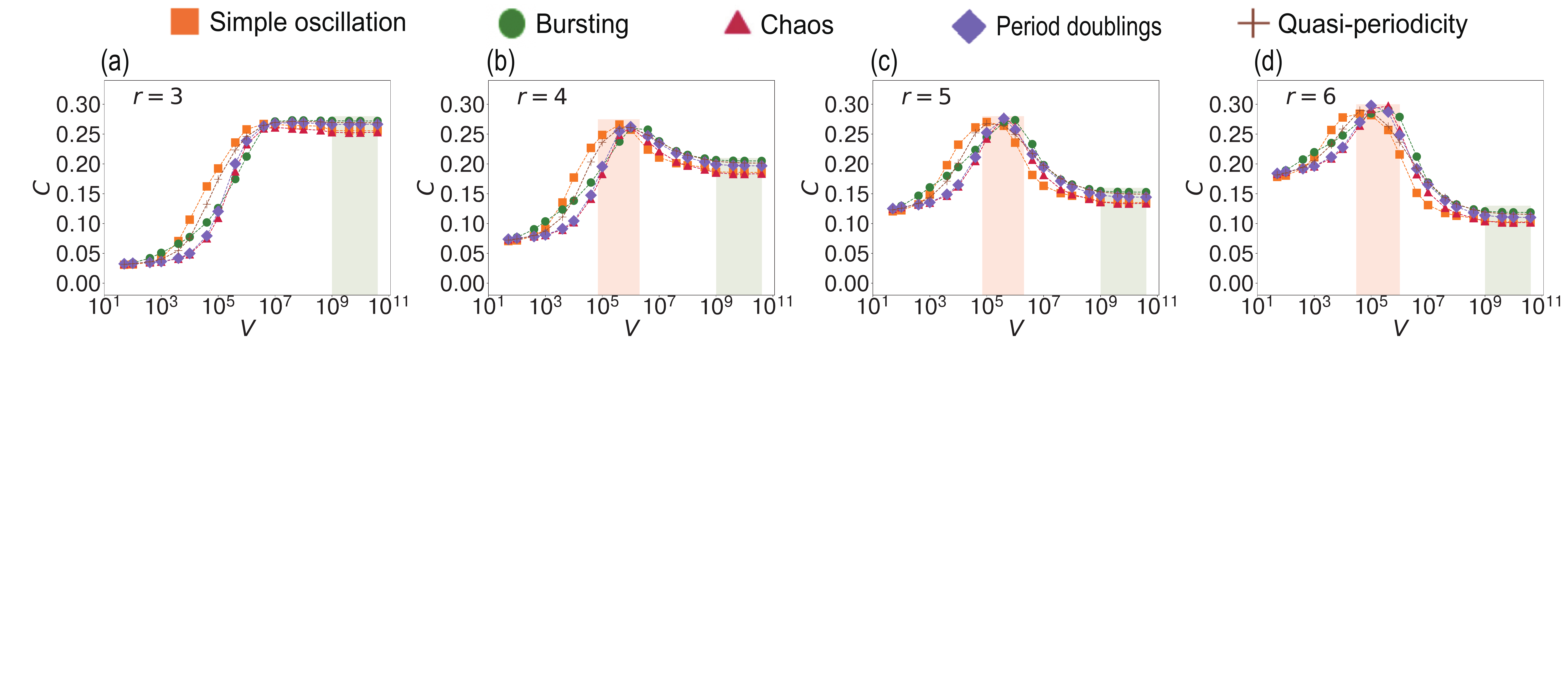} 
\vspace{-4cm}
\caption{(a--d) Statistical complexity measure $C$ against $V$ at $r=3,~4,~5,$ and $6$. The markers show the data for the five distinct cytosolic Ca$^{2+}$ states. The pink faded region indicates the region including the maximum values of $C$,  $C_{\mathrm{max}}$. The green faded region indicates the saturated values of $C$ in the large system size limit.}
\label{fig:ecplane1a}
\end{figure*}

We now investigate the interplay of intrinsic fluctuation and the complexity of various dynamic states of cytosolic Ca$^{2+}$ on the complexity-entropy (CH) causality plane for system size $V=10^6$, $10^5$, $10^4$ and $10^2$. 

In Fig.~\ref{fig:plane}, we plot the $(H,C)$ components of the six distinct Ca$^{2+}$ dynamics on the CH causality plane (where the embedding dimension is $r=6$). The solid lines represent the theoretical curves for the maximum and minimum of statistical complexity $C$~\cite{calbet2001tendency} (see Appendix~\ref{app:Ae}). 
The analysis shows the six dynamical states of cytosolic Ca$^{2+}$ as distinct states on the CH causality plane, and there exists a nontrivial relationship between the statistical complexity and the permutation entropy of cytosolic Ca$^{2+}$ dynamics. 

When $V=10^6$, all the oscillatory states are placed in the range of $0<H<0.5$ (see also Fig.~\ref{fig:ecplane}), suggesting relatively less disordered cytosolic Ca$^{2+}$ dynamics. As $V$ decreases to $10^5$, the ($H,C$) components of all the periodic and chaotic Ca$^{2+}$ states shift towards the center of the CH causality plane, with higher values of $C$ and $H$. This indicates that compared to $V=10^6$, the Ca$^{2+}$ oscillation system attains more complexity and information (entropy) driven by a larger intrinsic fluctuation~\cite{lopez1995statistical}. Upon further increasing intrinsic fluctuations ($V=10^2$), the dynamics of cytosolic Ca$^{2+}$ tend to demonstrate larger entropy ($H$) but reduced complexity ($C$).  Despite the increase in entropy ($H$), the complexity $C$ decreases with an increased intrinsic fluctuation due to a reduction in disequilibrium $DE$, as defined in Eq.~\eqref{eq:scm}. In the case of $V= 10^2$, intrinsic fluctuations destroy the ordered patterns in all the oscillatory states, signifying a transition towards disorder states. Importantly, the $V$-dependency shows that the statistical complexity of cytosolic Ca$^{2+}$ dynamics attains the maximum when the system has the intermediate size of $V\sim 10^5$. 

In Fig.~\ref{fig:plane}, additionally, we compare the complexity-entropy behavior of cytosolic Ca$^{2+}$ with that of fractional Brownian motion (fBm)~\cite{jeon2010fractional, rosso2007distinguishing}. The $(H,C)$ component of fBm is evaluated with the Hurst exponent from $\mathcal{H}=0.5$ to 0.9~\cite{rosso2007distinguishing}. As the comparison shows, intriguingly, the cytosolic Ca$^{2+}$ with $V=10^2$--$10^5$ exhibits a complexity-entropy structure akin to that of fBm. The $(H,C)$ states of Ca$^{2+}$ dynamics 
for $V=10^5$ (plus signs) nearly follow fBm process with $\mathcal{H}=0.9$ (a persistent Gaussian random walk~\cite{jeon2010fractional, rosso2007distinguishing}). For $V=10^2$, notably, the $(H,C)$ values of Ca$^{2+}$ dynamics closely align with that of fBm with $\mathcal{H}=0.5$, i.e., ordinary Brownian motion. 

We further investigate the $H$ and $C$ profiles for varying system size $V$.  
Figs.~\ref{fig:ecplane1}(a)--(d) show $H$ vs $V$ in log-log scale for given embedding dimensions of $r=3$, $4$, $5$, and $6$. We note that $H$ exhibits a power-law behavior in the intermediate range of $V\in [10^5,~10^7]$. The solid lines depict the best-fit power-law with $H(V)=aV^{-b}$ ($b$: the power-law exponent, $a$: a constant).  The fits have the goodness of fit $r^2$ score of $\approx 0.98$--$0.99$, and the corresponding values of $b$ are given in the legend. This power-law behavior can be interpreted in such a way that the cytosolic Ca$^{2+}$ dynamics of intermediate system size contain scale-free or fractal patterns in the dynamical structures~\cite{hong2006power,mehri2016power,nogueira2017exploring}. If $V\geq 10^9$, as observed, the system under negligible intrinsic fluctuations exhibits deterministic dynamics without disorder. Consequently, the measure of disorder, $H$, becomes a constant independent of $V$ in this regime.  

To corroborate the fractal pattern in the cytosolic Ca$^{2+}$ dynamics, we analyze their multifractal structure using the method of multifractal detrended fluctuation analysis (MFDFA)~\cite{kantelhardt2001detecting,kantelhardt2002multifractal}. This methodology generalizes the detrended fluctuation analysis (DFA) method~\cite{peng1994mosaic}, numerically identifying long-range correlations in non-stationary time series. Further descriptions of multifractality and the MFDFA algorithm are explained in Appendix~\ref{app:A2}. Here is the main ingredient of the MFDFA analysis. (1) For each dynamical state of cytosolic Ca$^{2+}$, we calculate the fluctuation function $F_q(s)$ [Eq.~\eqref{eq:fluc}] for varying scale length $s$ with a given moment $q$ ranging from $-10$ to $10$. (2) We define the generalized Hurst exponents $h(q)$ in the scaling relation of $F_q(s)\sim s^{h(q)}$ and estimate $h(q)$ from the fitting of $F_q(s)$. Fig.~\ref{fig:appendix_figure2}(a) shows $h(q)$ vs $q$ for $V=10^5$ (Left), $10^6$ (Middle), and $10^7$ (Right). Different colors represent different dynamic states, as indicated at the top of the figure. Notably, it is observed that $h(q)$ has a significant dependence on $q$, illustrating fractality or, more precisely, multifractality. The latter signifies the different nature of correlations in large and small fluctuations within the time series $x(t)$~\cite{kantelhardt2001detecting,kantelhardt2002multifractal}. Additionally, we calculate the classical multifractal scaling exponent $\tau(q)$ using Eq.~\eqref{eq:tau} and plot it against $q$ (Fig.~\ref{fig:appendix_figure2}(b)). The bi-linear nature of the $\tau(q)$ curves is observed, which indicates intrinsic multifractality arising from non-linear correlations~\cite{zhou2006inverse}. 

We also examine whether $h(q)$ for the various states of cytosolic Ca$^{2+}$ is consistent with that of fBm. 
Fig.~\ref{fig:fbm} presents the comparison between (a) Ca$^{2+}$ dynamics of $V=10^5$ and fBm of $\mathcal{H}=0.9$ (black dashed line) and between (b) Ca$^{2+}$ states of $V=10^2$ and fBm of $\mathcal{H}=0.5$. Since fBm is a monofractal process, as expected, $h(q)$ is evaluated to be a $q$-independent constant from our MFDFA analysis. Although the monofractal fBm cannot explain $h(q)$ of the Ca$^{2+}$ dynamics over the entire $q$ domain, it explains well the $h(q)$ with $q<0$, i.e., the short-time structures of the Ca$^{2+}$ dynamics for both cases of $V=10^2$ and $10^5$. This finding implicates that the small-scale or local structure of the Ca$^{2+}$ time series resembles the fBm of a specific Hurst index, resulting in the similarity in the CH causality plane. Note that the large-scale structure of the Ca$^{2+}$ dynamics, characterized by $h(q>0)$, evidently deviates from that of fBm, as the cytosolic Ca$^{2+}$ 
oscillates on a large-time scale.

In Fig.~\ref{fig:ecplane1a}(a)--(d), we plot the statistical complexity $C$ against $V$. For the embedding dimension $r=3$ [Fig.~\ref{fig:ecplane1a}(a)], the statistical complexities monotonically increase with $V$ and then become stationary. In the cases of larger embedding dimensions ($r=4$, 5, and 6), the statistical complexities have the maximum value (indicated by the shaded pink region) at an intermediate system size $V\sim 10^5$--$10^6$ and then decrease to reach stationary values (indicated by the shaded green region). The occurrence of the maximum complexity $C$ within an intermediate intrinsic fluctuation can be understood in the following. By the definition of Eq.~\eqref{eq:stat}, $C$ is the product of disequilibrium ($DE$) and the normalized permutation entropy ($H$) as $C=DE \times H$. The disequilibrium $DE$, which can be considered a variant of the Kullback-Leibler divergence from information theory, measures the extra bits required to code a sample in terms of the reference probability distribution. In Fig.~\ref{fig:oscillations}, the  Ca$^{2+}$ dynamics tends to be more regular (with negligible fluctuation) when $V$ becomes larger, while it becomes more noisy as $V$ becomes smaller. In the former case of large $V$, the  Ca$^{2+}$ dynamics approaches the deterministic limit, where determinism drives the system far from the reference equiprobable distribution of $x$ states, resulting in a large disequilibrium $DE$, whereas entropy $H$ is small. In the latter case of small $V$, the large intrinsic fluctuation averages out the distributions, resulting in equally probable patterns with $P \approx U$. Hence, in this case, $DE$ minimizes, while $H$ maximizes. Therefore, in these extreme conditions of $V$, the complexity $C$ is not maximized; it will attain the maximum within an intermediate range of $V$ where both $DE$ and $H$ have large values. An interesting finding in the literature is that the intracellular calcium oscillations have a maximum effective signal-to-noise ratio at a system size of $V \approx 10^6 ~\mu$m$^3$~\cite{li2005internal,zhu2007mesoscopic},  signifying the presence of an optimal system size or an optimal level of intrinsic fluctuation in real systems~\cite{li2005internall}. 

\begin{figure}
\centering
\includegraphics[scale=0.4]{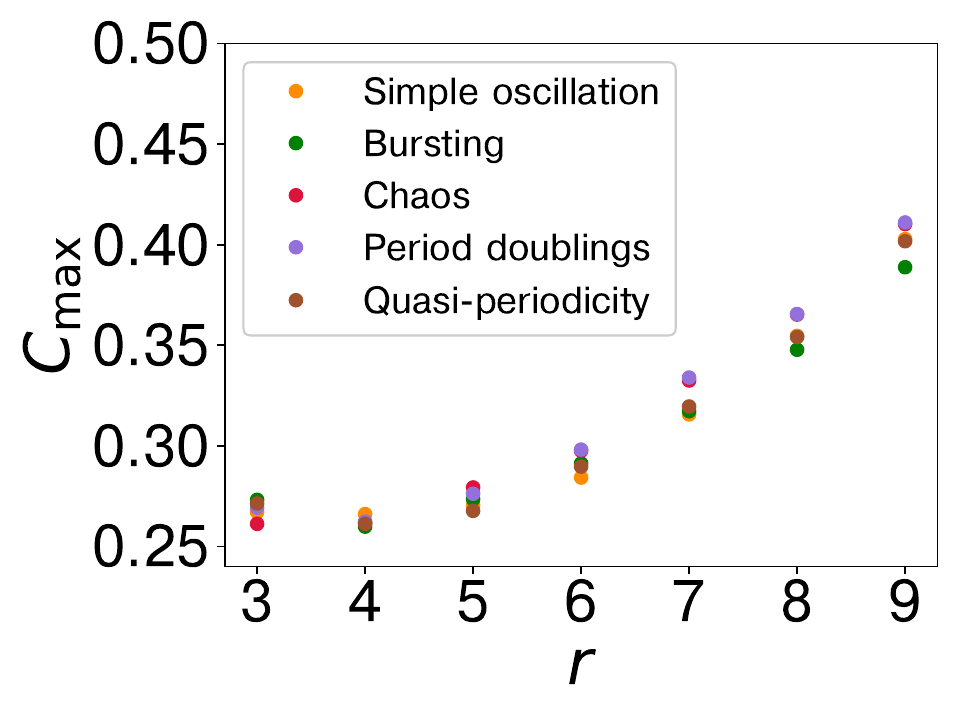}
\caption{Dependence of the maximum value of the statistical complexity measure, $C_{\mathrm{max}}$ on the embedding dimension $r$.}
    \label{fig:ecplane2}
\end{figure}

Finally, we examine the effect of the embedding dimension $r$ on the statistical complexity measure $C_\mathrm{max}$. Fig.~\ref{fig:ecplane2} shows $C_\mathrm{max}$ for all the oscillatory cytosolic Ca$^{2+}$ states for increasing  $r$ up to the largest value that can be investigated within the total observation time of simulated time series. We find that, in the explored range of $r$, $C_\mathrm{max}$ occurs within the intermediate system size and monotonically increases with $r$. The latter behavior emerges due to the fact that the two components of $C$, i.e., permutation entropy and disequilibrium, increase with the embedding dimension (or permutation order) $r$, as defined in Eq.~\eqref{eq:scm}.     
\section{conclusions}
\label{sec:five}
The collective behavior of a network (e.g., chemical reaction network) of diverse components (e.g., molecular species) within a complex system often leads to a variety of complex dynamic states~\cite{boccaletti2006complex, eskov2016evolution}, each playing a significant functional role. In dynamical systems, a steady state typically signifies an equilibrium state~\cite{strogartz1994nonlinear}, oscillations reflect the active state of the system~\cite{dolmetsch1998calcium}, and chaos is known to play a crucial role in information processing~\cite{korn2003there,riaz2008chaotic}. Experimental observations indicate that calcium ions (Ca$^{2+}$) within biological cells exhibit a variety of complex dynamic behaviors, including complex oscillations. The Ca$^{2+}$ oscillations show stochastic dynamics and a critical determinant factor for the dynamics is the system size $V$ of the biological cell. $V$ modulates the intrinsic fluctuations in Ca$^{2+}$ dynamics such that intrinsic fluctuation scales as $\sim \frac{1}{\sqrt{V}}$. A systematic exploration of the interplay between intrinsic fluctuation and the complexity of observed dynamics of intracellular Ca$^{2+}$ has not been thoroughly addressed. Our study addresses this gap through a comprehensive analysis of the complexities associated with the different dynamics of cytosolic Ca$^{2+}$ and their interplay with intrinsic fluctuations. Leveraging complexity measures such as permutation entropy and statistical complexity, our results unravel intricate relationships between intrinsic fluctuations and the complexity of Ca$^{2+}$ dynamics across various states. 

We perform stochastic simulations employing the chemical Langevin equation (CLE) to model the complex dynamics of intracellular Ca$^{2+}$. By varying the system size $V$, we delve into the interplay between intrinsic fluctuation and the complexity of cytosolic Ca$^{2+}$ dynamics. Strong intrinsic fluctuations lead to the breakdown of the ordered states in cytosolic Ca$^{2+}$ concentration. We find that permutation entropy effectively characterizes the complexities of the different dynamic states and their changes due to intrinsic fluctuation. With permutation entropy, chaos is found to be highly sensitive to intrinsic fluctuation. The complexity-entropy (CH) causality plane, initially proposed for distinguishing noise and chaos, proved valuable for assessing the complexities of diverse, dynamic states and analyzing their interplay with intrinsic fluctuation. The distinct dynamic states of cytosolic Ca$^{2+}$ exhibit varying positions within the theoretical bounds of the CH causality plane as intrinsic fluctuation varies, indicating varying degrees of complexity and information content. In an intermediate range of $V$ (or intrinsic fluctuation), the normalized permutation entropy $H$ follows a power-law behavior, suggesting the presence of scale-free or fractal patterns in the dynamical structures of cytosolic Ca$^{2+}$, furthermore corroborated by the multifractal detrended fluctuation analysis (MFDFA), providing additional insights into the nature of correlations of fluctuations in the time series of cytosolic Ca$^{2+}$ concentration. Additionally, we observe peak values in the statistical complexities of the periodic and chaotic states of cytosolic Ca$^{2+}$ at an intermediate level of intrinsic fluctuation when adjusting the permutation order or embedding dimension $r$. This intermediate range is characterized by fractal patterns with permutation entropy analysis. Such high-complexity states may correspond to optimal Ca$^{2+}$ dynamics, holding potential biological significance, for instance, information transfer within signaling pathways. This study deepens our understanding of how intrinsic fluctuations dynamically regulate the complex behaviors of intracellular Ca$^{2+}$ across diverse states. Our study is intimately related to understanding the influence of intrinsic fluctuation on the dynamics of several biological systems at microscopic and mesoscopic scales. Investigating the effect of intrinsic fluctuations on the complex dynamics of non-linear biochemical systems remains a topic of fundamental importance. 

Oscillations are well-established far-from-equilibrium phenomena~\cite{prigogine1978time}. For a non-equilibrium system with oscillatory dynamics, the stochastic process along the oscillating trajectory is dynamically irreversible~\cite{seara2021irreversibility} when transitioning from an initial state $x_i$ to some final state $x_f$ and then from $x_f$ to $x_i$. The energy dissipation in such a non-equilibrium system can be quantitatively measured by estimating the total entropy production~\cite{seifert2004fluctuation}, denoted as $\Delta\Sigma_{\mathrm{tot}}$, in the trajectory. $\Delta\Sigma_{\mathrm{tot}}$ can be calculated from the probabilities of the forward and backward paths, given by $p[x(t)|x_i]$ and $p[x(t)|x_f]$, respectively, such that $\Delta\Sigma_{\mathrm{tot}}=\ln\frac{p[x(t)|x_i]}{p[x(t)|x_f]}+\ln\frac{p_0(x_i)}{p_0(x_f)}$, with normalized distributions $p_0(x_i)$ and $p_0(x_f)$~\cite{andrieux2007entropy,xiao2009stochastic}. Estimating entropy production in stochastic trajectories from experiments or simulations is an active area of research bearing fundamental importance for a deeper understanding of non-equilibrium fluctuations and non-equilibrium properties of dynamical systems. It will be interesting to analyze entropy production in the non-equilibrium system of intracellular calcium oscillations and will be carried out in future work. 
\section*{Acknowledgements}
\noindent This research is supported by ALC's appointment to the YST program at the APCTP through the Science and Technology Promotion Fund and Lottery Fund of the Korean Government (and local governments of Gyeongsangbuk-do Province and Pohang city) and the National Research Foundation of Korea, Grant No.~RS-2023-00218927 and 2021R1A6A1A10042944 (JHJ). ALC acknowledges Dr. Mohammad Zubbair Malik (Scientist, Dasman Diabetes Institute, Kuwait) for the initial discussions. ALC expresses a special thanks to Yeongjin Kim for his kind help in preparing the figures. \\
\section*{Conflict of Interests }
\noindent The authors have no conflicts to disclose.
\section*{Data Availability Statement}
\noindent The data that support the findings of this study are available within the article itself.\\
\appendix
\section{}
\label{app:A1}
\renewcommand{\thefigure}{A.\arabic{figure}}
\setcounter{figure}{0}
\begin{table*}
\centering
\caption{Ordinary differential equations describing the intracellular calcium oscillation model~\eqref{eq:model} with the corresponding reaction channels. The concentrations of the cytosolic Ca$^{2+}$, Ca$^{2+}$ stored and InsP$_3$ are represented by $x(=X/V)$, $y(=Y/V)$, and $z(=Z/V)$, respectively, with $V$ as the system size.}
\label{table:one}
\begin{tabular}{|p{5.5cm}|p{3.3cm}|p{3.3cm}|}
\hline
Ordinary differential equations & Transition of states  & Propensity functions   \\
\hline
$\frac{dx}{dt}=V_0+V_1\beta-V_2+V_3+k_fy-kx$ &\ce{$X \rightarrow X + 1$} &  $V V_0$\\
&\ce{$X \rightarrow X + 1$} &  $V V_1\beta $\\
&\ce{$X \rightarrow X - 1$} &  $V V_2$\\
&\ce{$X \rightarrow X + 1$} &  $V V_3$\\
&\ce{$X \rightarrow X + 1$} &  $V k_fy$\\
&\ce{$X \rightarrow X - 1$} &  $V kx$\\
\hline
$\frac{dy}{dt}=V_2-V_3-k_fy$ &\ce{$Y \rightarrow Y + 1$} & $V V_2$\\
&\ce{$Y \rightarrow Y - 1$} & $V V_3$\\
&\ce{$Y \rightarrow Y - 1$} & $V k_fy$\\
\hline 
$\frac{dz}{dt}=\beta V_4-V_5-\epsilon z$ &\ce{$Z \rightarrow Z + 1$} &$V \beta V_4$\\ 
&\ce{$Z \rightarrow Z - 1$} &$V V_5$\\
&\ce{$Z \rightarrow Z - 1$} &$V \epsilon z$\\
\hline
\end{tabular}
\end{table*}
\begin{table*}
\centering
\caption{Parameter values used in the numerical simulation of the intracellular calcium oscillation model~\eqref{eq:model} \cite{houart1999bursting}}
\label{table:two}
\begin{tabular}{ |p{2.6cm}|p{2.3cm}|p{2.2cm}|p{2.3cm}|p{2.3cm}|p{2.5cm}|p{2.3cm}|}
\hline
Parameters & Steady state & Simple   & Bursting & Chaos &Period doubling  &Quasi- \\
  & & periodic &  & &sequences & periodicity\\
   & & oscillation  &  & & & \\
\hline
$V_0$ ($\mu$M min$^{-1}$) & 2& 2 &$2$ &2 &2 &2\\
$V_{1}$ ($\mu$M min$^{-1}$) & 2 & 2     & $2$ &2 &2 &2\\
$\beta$ & 0.01 &0.5 & $0.46$ &0.65 &0.7 &0.51\\
$V_{M2}$ ($\mu$M min$^{-1}$)  &6 & 6 & $6$ &6 &6 &6\\
$k_2$ ($\mu$M) & 0.1 & 0.1 & $0.1$ &0.1 &0.1 &0.1\\
$V_{M3}$($\mu$M min$^{-1}$) & 20 & 20 & $20$ & 30 &30 & $20$    \\
$k_x$ ($\mu$M) & 0.5 & 0.5 & $0.3$  & 0.6 &0.6 &$0.5$ \\
$k_{y}$ ($\mu$M) & 0.2 & 0.2   & $0.2$ & 0.3  &0.3 & $0.2$ \\
$k_z$ ($\mu$M)  & 0.2 & $0.2$ & $0.1$  & $0.1$ &0.1 &$0.2$\\
$V_{M5}$($\mu$M min$^{-1}$)&$30$ & $5$ &$30$  & $50$ &50 &$30$\\
$k_{5}$ ($\mu$M) &$0.3$ & $1$     & $1$ & $0.3194$     &0.3194 &$0.3$\\
$k_d$ ($\mu$M) &$0.5$ &$0.4$ & $0.6$ &$1$ &1 &$0.5$ \\
$k_{f}$ (min$^{-1}$)   & $1$ & $1$ & $1$ & $1$ &1 &$1$\\
$k $ (min$^{-1}$) & $10$ & $10$ & $10$ & $10$ &10 &$10$   \\
$\epsilon $ (min$^{-1}$)& $0.1$ & $0.1$ & $0.1$ & $13$ &13 & $0.1$\\
$V_{4}$($\mu$M min$^{-1}$) & $5$ & $2$   & $2.5$ & $3$   &3 &$5$\\
$m$  & 2 & 2 & $4$ & 2 &2 &$2$\\
$p$   & 2 & 2 & $1$ & 1 &1 &$2$\\
$n$   & 4 & 4 & $2$ & 4 &4 &$4$\\
\hline
\end{tabular}
\end{table*}
\begin{figure}
\includegraphics[scale=0.43]{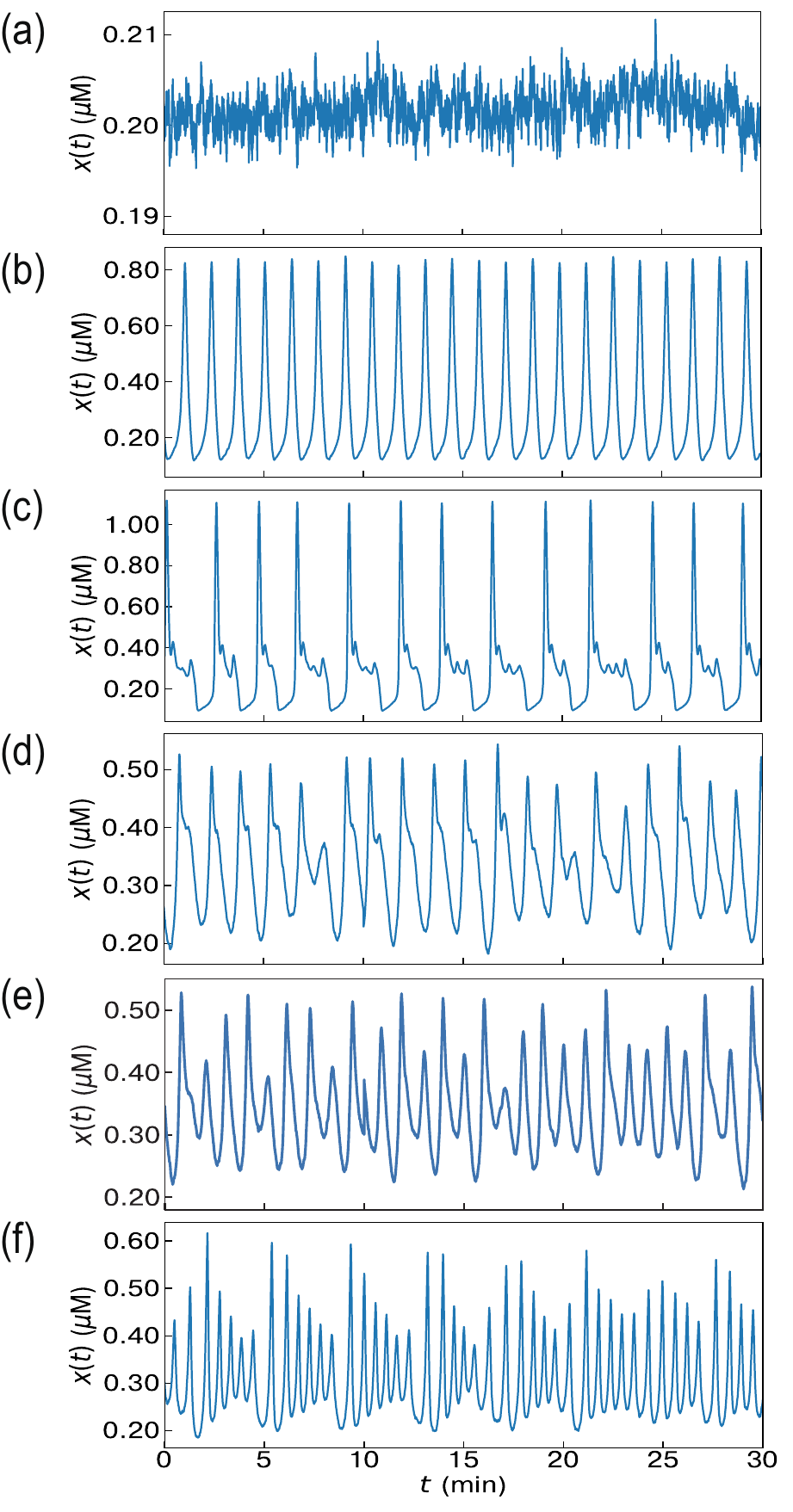}
    \caption{The time evolution of the cytosolic Ca$^{2+}$ concentration $x(t)$ for the different dynamic states: (a) steady state, (b) simple periodic oscillation, (c) bursting, (d) chaos, (e) period doubling sequences, and (f) quasi-periodicity, obtained by solving the chemical Langevin equation (CLE)~\eqref{eq:cle} at $V=10^5$. }
    \label{fig:appendix_figure0}
\end{figure}
\begin{figure}
\includegraphics[scale=0.4]{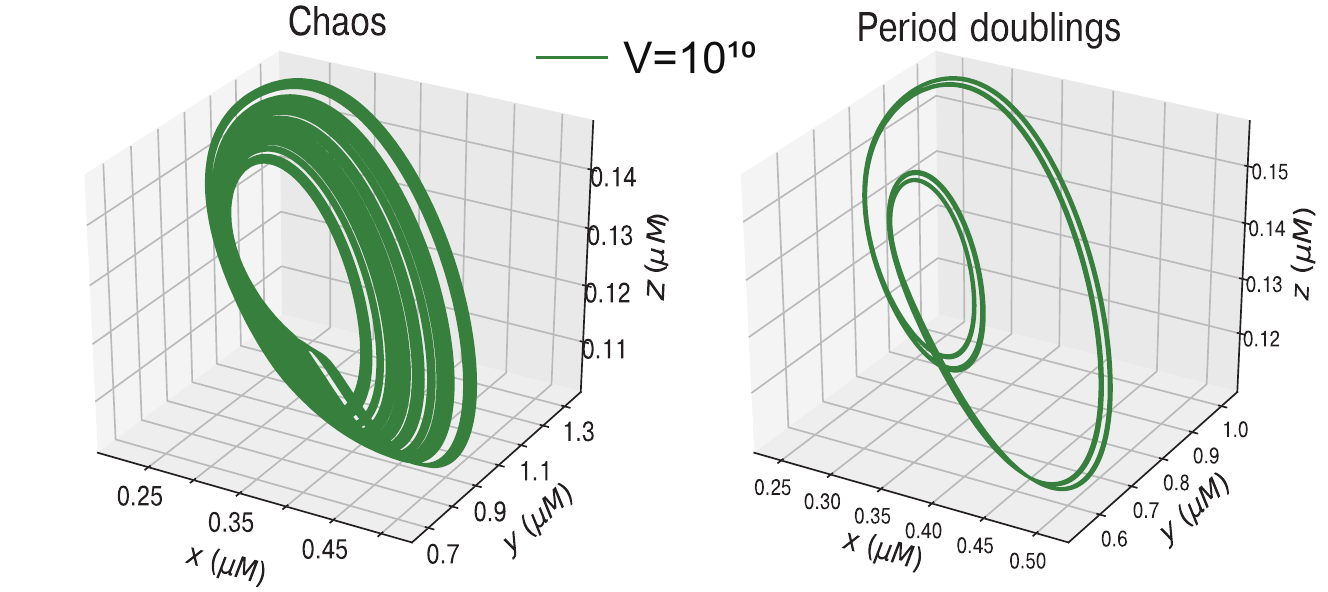} 
    \caption{Phase space plots of chaos and period-doubling of the intracellular Ca$^{2+}$ dynamics, obtained by solving the chemical Langevin equation (CLE)~\eqref{eq:cle} for $V=10^{10}$. The period doubling state is, in fact, understood as a sequence of two periodic doublings. }
    \label{fig:append}
\end{figure}
\begin{figure}
\includegraphics[scale=0.5]{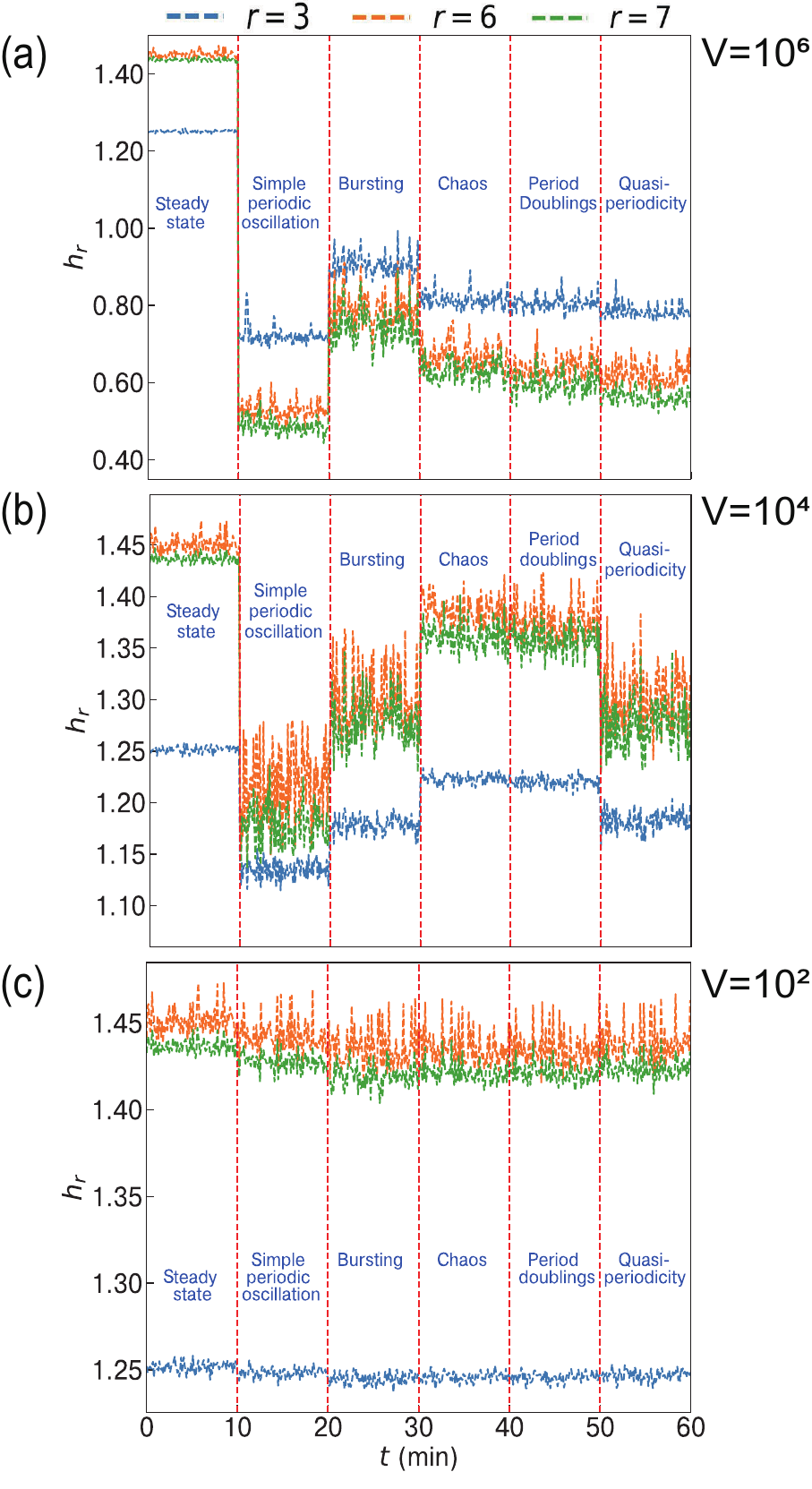} 
\caption{Time evolution of the permutation entropy per symbol of order $r$, denoted as $h_r$, for various dynamic states of cytosolic Ca$^{2+}$: (a) $V=10^6$, (b) $V=10^4$, and (c) $V=10^2$. To plot the time evolution of $h_r$, we divide the one-dimensional trajectory $x(t)$ of length $10^4$ into a number of non-overlapping smaller segments of equal size. Using a sliding window of size $r=3$ (blue), $6$ (orange), and $7$ (green), we then calculate the $h_r$ in each non-overlapping segment and obtain $h_r(t)$ over the total length of the trajectory. The $h_r(t)$ effectively captures the complexities of the fine temporal patterns in different dynamic states of cytosolic Ca$^{2+}$ concentrations and their interplay with intrinsic fluctuations. Vertical red lines mark the distinct dynamic states.}
    \label{fig:pe}
\end{figure}
In Table~\ref{table:one}, we perform stochastic modeling of the coupled, non-linear ordinary differential equations (ODEs) of the intracellular calcium oscillation model~\eqref{eq:model}. We translate the ODEs into a set of chemical reaction channels that show the transition of states of the $X, Y$, and $Z$ populations. Table~\ref{table:two} presents the values of the parameters used to solve the chemical Langevin equation (CLE)~\eqref{eq:cle} of the intracellular calcium oscillation. The values are taken from Ref.~\cite{houart1999bursting}.
\
\section{}
\label{app:Ae}
\renewcommand{\thefigure}{B.\arabic{figure}}
\setcounter{figure}{0}
\begin{figure}
\includegraphics[scale=0.3]{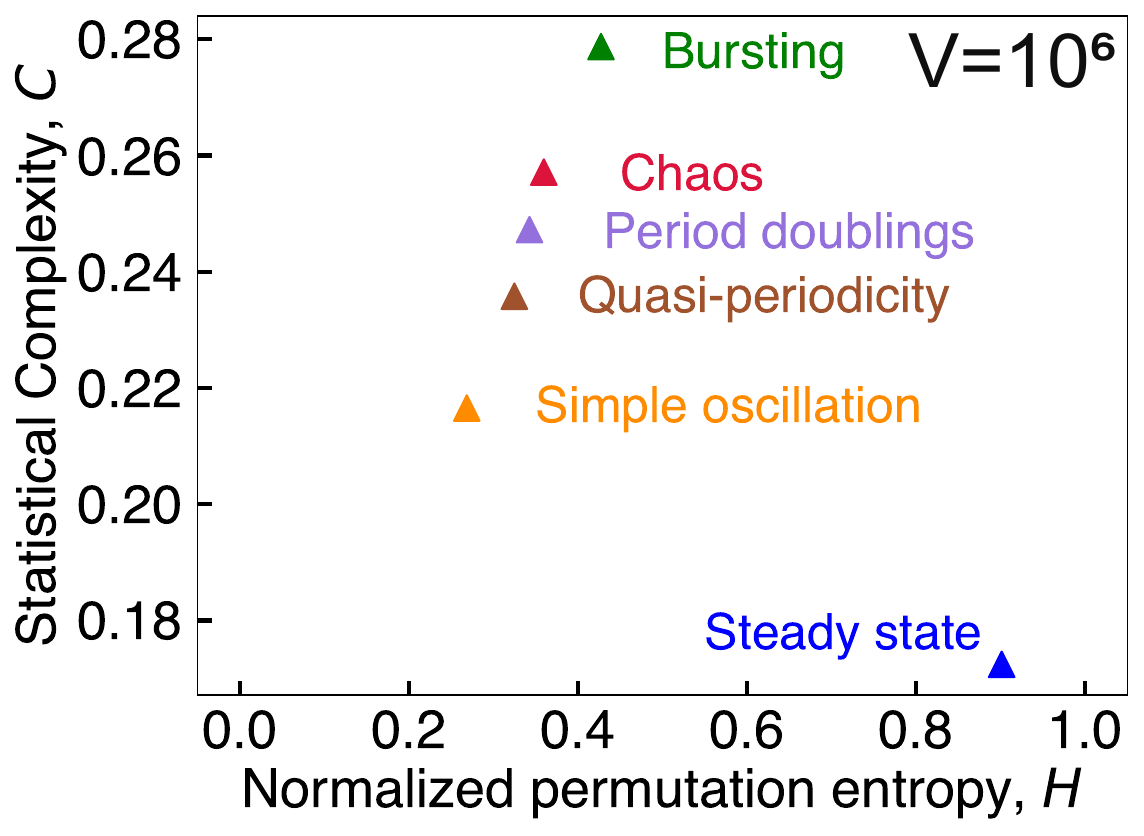}
    \caption{Location of the different dynamic states of the cytosolic Ca$^{2+}$ on the complexity-entropy (CH) causality plane at system size $V=10^6$.}
    \label{fig:ecplane}
\end{figure}
For a given normalized permutation entropy $H$, there exists a set of statistical complexity $C$ values between $C_\mathrm{min}$ and $C_\mathrm{max}$~\cite{rosso2007distinguishing}.  Here we briefly explain how to calculate $C_\mathrm{min}$ and $C_\mathrm{max}$ as described in Ref.~\cite{calbet2001tendency}.  Suppose a system has $M(=r!)$ possible accessible states $\{x_i \ ; \ i=1,2,\dots,M\}$ at a given scale with the probability $g_i$ of being in state $i$. At equilibrium, all states are equiprobable with $g_\mathrm{eq}=1/M$. For a given $M$,  a set of distributions giving $C_\mathrm{max}$ can be $\{g_1,g_i\}$ with $g_1=g_\mathrm{max}$ and $g_i=\frac{1-g_\mathrm{max}}{M-1}$.  The index $i=2,3,\dots,M$, and $g_\mathrm{max}$ runs from $1/M$ to 1.  Similarly,  a set of distributions giving $C_\mathrm{min}$ can be $\{g_1,g_i\}$ with $g_1=g_\mathrm{min}$ and $g_i=\frac{1-g_\mathrm{min}}{M-k-1}$, where $g_\mathrm{min}$ runs from 0 to $1/(M-k)$ with  $k=0,1,\dots,M-2$.

\section{Multifractality and multifractal detrended fluctuation analysis (MFDFA)}
\label{app:A2}
\renewcommand{\thefigure}{C.\arabic{figure}}
\setcounter{figure}{0}
\begin{figure}
\includegraphics[scale=0.16]{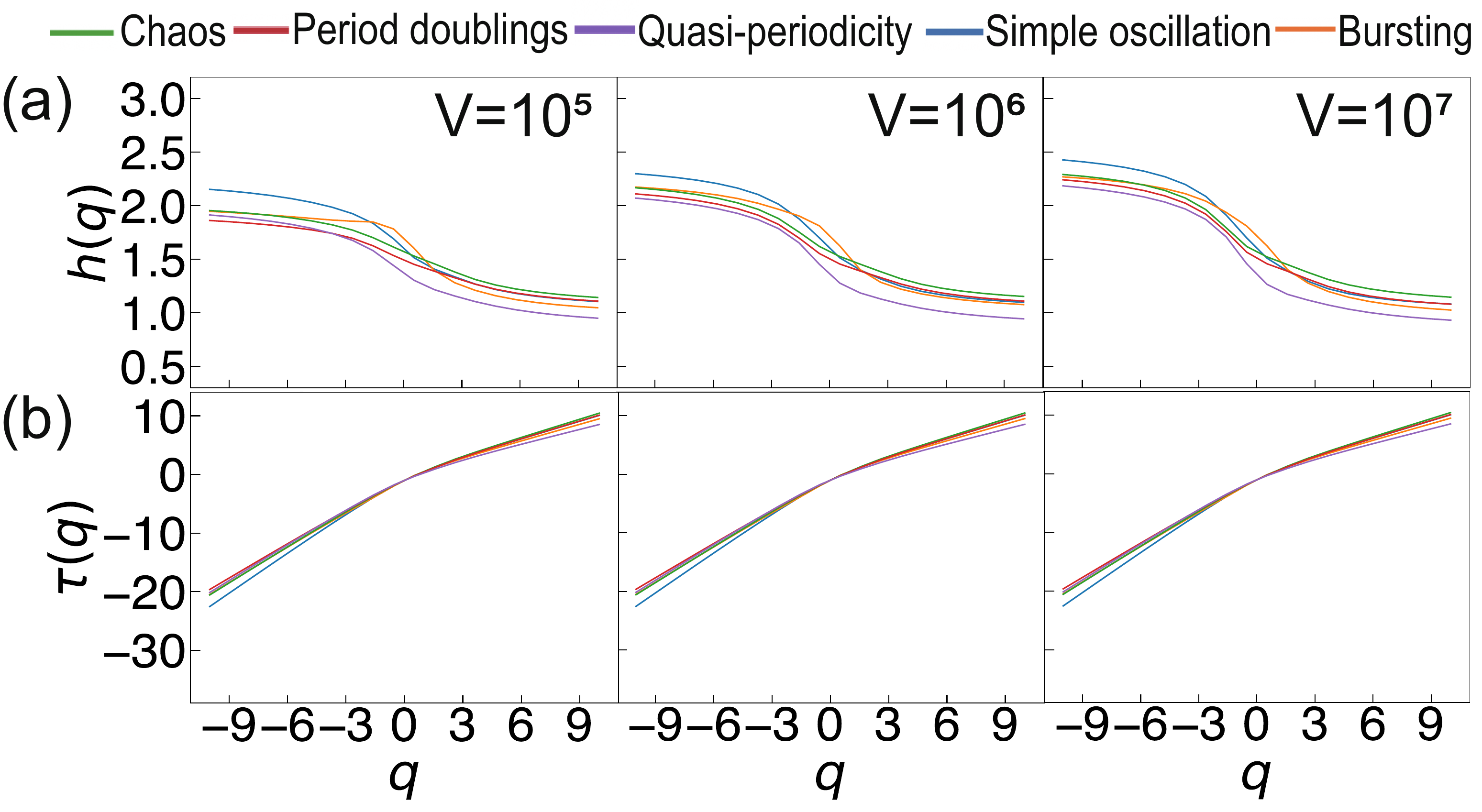}
    \caption{Plots of the multifractal parameters, namely (a) generalized Hurst exponent $h(q)$, and (b) classical multifractal scaling exponent $\tau(q)$ at the system size $V=10^5$ (Left), $V=10^6$ (Middle) and $V=10^7$ (Right). The different dynamical states are depicted by distinct colored curves, as seen in legend.}
    \label{fig:appendix_figure2}
\end{figure}
A self-affine process is a random process $z(t)$ that follows the scaling relation~\cite{RePEc:cwl:cwldpp:1164}
\begin{align}
	\frac{z(at)}{z(t)}=a^{C} \ ; \ \forall \ a>0, \label{eq:selfaffine1}
\end{align}
with scale factor $a$. The scaling exponent $C>0$ represents the fractal or self-similarity dimension. Fluctuations observed in the data generated by complex systems often obey a scaling relation of the type \eqref{eq:selfaffine1} on a broad range of scales~\cite{kantelhardt2002multifractal}. The theory of multifractals concerns a general scaling relation:  
\begin{equation}
	\frac{z(at)}{z(t)}= \Lambda(a)\ ; \forall \ t, \ 0<a. \label{eq:selfaffine2}
\end{equation}
Here, $z(t)$ and $\Lambda(a)$ are independent random functions. If the random process $z(t)$ is multifractal, then $\Lambda(a)$ satisfies~\cite{RePEc:cwl:cwldpp:1164}
\begin{align}
	\Lambda(a_1a_2\dots a_m)&=\Lambda_1(a_1)\ \Lambda_2(a_2)\dots \Lambda_n(a_m) \ ;  \nonumber \\
	& \ 0 <a_1,a_2,\dots,a_m\leq 1, \label{eq:selfaffine3}
\end{align}
where $\Lambda_{1,2, \dots,m}$ are $m$ independent copies of $\Lambda$ for various local scales $a_{1,2, \dots,m}$.
Using Eqs.~\eqref{eq:selfaffine1}$\&$\eqref{eq:selfaffine2}, each local scale $a_{\kappa}$ of Eq.~\eqref{eq:selfaffine3} has the local fractal dimension $C_{\kappa}$, where the local scaling function follows 
\begin{equation}
	\Lambda_{\kappa}(a_{\kappa})\sim a_{\kappa}^{C_{\kappa}}. \label{eq:selfaffine4}
\end{equation} 
If $a_1=a_2=\dots =a_m=a$ in Eq.~\eqref{eq:selfaffine3}, then the multifractal process becomes a monofractal. Using Eq.~\eqref{eq:selfaffine4} in~\eqref{eq:selfaffine3}, $\Lambda=a^{C_1+C_2+\dots+C_m}=a^C$, where $C=C_1+C_2+...+C_n$. Therefore, for a uni-fractal system, the scaling function $\Lambda$ characterizes a homogenous fractal structure or behavior with a single scaling exponent $C$ at all scales $a$. On the other hand, multifractality can characterize richer, heterogenous structures in complex systems with a set of scaling exponents corresponding to local spatial or temporal patterns.

Given a time series $x_k$ of length $N$, we can investigate the multifractality of the series using the multifractal detrended fluctuation analysis (MFDFA) algorithm. We describe the algorithm and the associated physical interpretations as follows.
\begin{enumerate}
\item Calculate the profile of the time series as
\begin{equation}
Y(i) \equiv \sum_{k=1}^{i} (x_k-\langle x \rangle), \ \ i=1,\dots,N. \label{eq:step1}
\end{equation}
\item Divide the profile $Y(i)$ into $N_s \equiv \mathrm{int}\left(\frac{N}{s}\right)$ non-overlapping segments of equal scale length $s$. If $N$ is not a multiple of $s$, the same procedure is repeated from the end resulting in $2N_s$ segments.
\item Calculate the trend of each $2N_s$ segment using the least-square fitting method. The variance is 
\begin{equation}
F^2(\nu,s)\equiv \frac{1}{s} \sum_{i=1}^{s} \{Y[(\nu-1)s+i]-y_{\nu}(i)\}^2,
\end{equation}
for each segment $\nu=1,...,N_s$, and
\begin{equation}
F^2(\nu,s)\equiv \frac{1}{s} \sum_{i=1}^{s} \{Y[(N-(\nu-N_s)s+i]-y_{\nu}(i)\}^2\nonumber
\end{equation}
for $\nu=N_s+1,...,2N_s$. The $y_\nu(i)$ is the fitting polynomial in segment $\nu$.
\item Determine the $q^{\mathrm{th}}$ order fluctuation function $F_q(s)$ by averaging over all the segments as
\begin{equation}
F_q(s)\equiv \left \{ \frac{1}{2N_s}\sum_{\nu=1}^{2Ns} [F^2(\nu,s)]^{q/2}\right \}^{1/q}. \label{eq:fluc}
\end{equation}
Calculate $F_q(s)$ for different time scales $s$ and $q$.
\item If the time series $x_k$ is long-range power-law correlated, $F_q(s)$ follows a power-law with the scale length $s$ as,
\begin{equation}
F_q(s) \sim s^{h(q)}, \label{eq:pl}
\end{equation}
where the power-law exponent $h(q)$ is known as the generalized Hurst exponent. In the log-log plots of $F_q(s)$ versus $s$ for different values of $q$, the exponent $h(q)$ corresponds to the slopes of the graphs. For $q=0$, the fluctuation function $F_q(s)$ in Eq.~\eqref{eq:fluc} diverges and hence, we use the expression
\begin{equation}
F_0(s)\equiv \exp \left \{ \frac{1}{4N_s}\sum_{\nu=1}^{2N_s} \ln[F^2(\nu,s)]\right \} \sim s^{h(0)}.
\end{equation}
\end{enumerate}
If $h(q)$ is a constant independent of $q$, the time series has a monofractal structure. If the characteristics of small- and large-scale fluctuations differ, there is a significant dependence of $h(q)$ on $q$, indicating multifractal behavior. For $q>0$, $h(q)$ describes the scaling behavior of segments with large fluctuations. For $q<0$, $h(q)$ describes the scaling behavior of segments with small fluctuations. While $q>0$ accounts for large-scale or global patterns in time series, $q<0$ accounts for small-scale or local patterns. The generalized exponent $h(q)$ is related to the classical multifractal scaling exponent $\tau(q)$ from the standard partition function-based multifractal formalism through the relation,
\begin{equation}
\tau(q) = qh(q)-1. \label{eq:tau}
\end{equation}
If $h(q)$ has a monofractal behavior, $\tau(q)$ is linear. A non-linear $\tau(q)$ thus indicates multifractal behavior.
\bibliographystyle{refstyl}
\bibliography{references}

\providecommand{\href}[2]{#2}\begingroup\raggedright\begin{thebibliography}{100}

\bibitem{mitchell2009complexity}
M.~Mitchell, {Complexity: A guided tour}, Oxford university press (2009).

\bibitem{strogartz1994nonlinear}
S.H.~Strogartz, {Nonlinear dynamics and chaos: With applications to physics,
  biology},  {Chemistry and Engineering} {\bfseries 441} (1994).

\bibitem{nicolis1989exploring}
G.~Nicolis and I.~Prigogine, {Exploring complexity: An introduction}, .

\bibitem{nicolis2012foundations}
G.~Nicolis and C.~Nicolis, {Foundations of complex systems: emergence,
  information and predicition}, World Scientific (2012).

\bibitem{pisarchik2014control}
A.N.~Pisarchik and U.~Feudel, {Control of multistability},  {Physics Reports}
  {\bfseries 540} (2014) 167.

\bibitem{rickles2007simple}
D.~Rickles, P.~Hawe and A.~Shiell, {A simple guide to chaos and complexity},
  {Journal of Epidemiology \& Community Health} {\bfseries 61} (2007) 933.

\bibitem{aguirre2009fractal}
J.~Aguirre, R.L.~Viana and M.A.~Sanju{\'a}n, {Fractal structures in nonlinear
  dynamics},  {Reviews of Modern Physics} {\bfseries 81} (2009) 333.

\bibitem{bressloff2014stochastic}
P.C.~Bressloff, {Stochastic processes in cell biology}, vol.~41, Springer
  (2014).

\bibitem{allen1994evaluation}
C.~Allen and C.F.~Stevens, {An evaluation of causes for unreliability of
  synaptic transmission.},  {PNAS} {\bfseries 91} (1994) 10380.

\bibitem{shahrezaei2008stochastic}
V.~Shahrezaei and P.S.~Swain, {The stochastic nature of biochemical networks},
  {Current opinion in biotechnology} {\bfseries 19} (2008) 369.

\bibitem{blossey2006compositional}
R.~Blossey, L.~Cardelli and A.~Phillips, {A compositional approach to the
  stochastic dynamics of gene networks},  in {Transactions on Computational
  Systems Biology IV}, pp.~99--122, Springer (2006).

\bibitem{kolomeisky2013motor}
A.B.~Kolomeisky, {Motor proteins and molecular motors: how to operate machines
  at the nanoscale},  {Journal of Physics: Condensed Matter} {\bfseries 25}
  (2013) 463101.

\bibitem{qian2007phosphorylation}
H.~Qian, {Phosphorylation energy hypothesis: open chemical systems and their
  biological functions},  {Annu. Rev. Phys. Chem.} {\bfseries 58} (2007) 113.

\bibitem{qian2013stochastic}
H.~Qian, {Stochastic physics, complex systems and biology},  {Quantitative
  Biology} {\bfseries 1} (2013) 50.

\bibitem{prigogine1978time}
I.~Prigogine, {Time, structure, and fluctuations},  {Science} {\bfseries 201}
  (1978) 777.

\bibitem{epstein1998introduction}
I.R.~Epstein and J.A.~Pojman, {An introduction to nonlinear chemical dynamics:
  oscillations, waves, patterns, and chaos}, Oxford university press (1998).

\bibitem{goldbeter2018dissipative}
A.~Goldbeter, {Dissipative structures in biological systems: bistability,
  oscillations, spatial patterns and waves},  {Philos. Trans. R. Soc. A}
  {\bfseries 376} (2018) 20170376.

\bibitem{chung2022thermodynamics}
B.J.~Chung, B.~De~Bari, J.~Dixon, D.~Kondepudi, J.~Pateras and A.~Vaidya, {On
  the thermodynamics of self-organization in dissipative systems: Reflections
  on the unification of physics and biology},  {Fluids} {\bfseries 7} (2022)
  141.

\bibitem{poon1995controlling}
L.~Poon and C.~Grebogi, {Controlling complexity},  {Physical review letters}
  {\bfseries 75} (1995) 4023.

\bibitem{hilfinger2011separating}
A.~Hilfinger and J.~Paulsson, {Separating intrinsic from extrinsic fluctuations
  in dynamic biological systems},  {PNAS} {\bfseries 108} (2011) 12167.

\bibitem{rao2002control}
C.V.~Rao, D.M.~Wolf and A.P.~Arkin, {Control, exploitation and tolerance of
  intracellular noise},  {Nature} {\bfseries 420} (2002) 231.

\bibitem{kaern2005stochasticity}
M.~Kaern, T.C.~Elston, W.J.~Blake and J.J.~Collins, {Stochasticity in gene
  expression: from theories to phenotypes},  {Nature Reviews Genetics}
  {\bfseries 6} (2005) 451.

\bibitem{samoilov2006fluctuations}
M.S.~Samoilov, G.~Price and A.P.~Arkin, {From fluctuations to phenotypes: the
  physiology of noise},  {Science's STKE} {\bfseries 2006} (2006) re17.

\bibitem{gillespie2007stochastic}
D.T.~Gillespie, {Stochastic simulation of chemical kinetics},  {Annu. Rev.
  Phys. Chem.} {\bfseries 58} (2007) 35.

\bibitem{gillespie2000chemical}
D.T.~Gillespie, {The chemical langevin equation},  {The Journal of Chemical
  Physics} {\bfseries 113} (2000) 297.

\bibitem{ye2007growing}
B.~Ye, Y.~Zhang, W.~Song, S.H.~Younger, L.Y.~Jan and Y.N.~Jan, {Growing
  dendrites and axons differ in their reliance on the secretory pathway},
  {Cell} {\bfseries 130} (2007) 717.

\bibitem{canham1968distribution}
P.~Canham and A.C.~Burton, {Distribution of size and shape in populations of
  normal human red cells},  {Circulation research} {\bfseries 22} (1968) 405.

\bibitem{shen2021tcf21+}
Y.-c.~Shen, A.N.~Shami, L.~Moritz, H.~Larose, G.L.~Manske, Q.~Ma et~al.,
  {Tcf21+ mesenchymal cells contribute to testis somatic cell development,
  homeostasis, and regeneration in mice},  {Nature communications} {\bfseries
  12} (2021) 3876.

\bibitem{balazsi2011cellular}
G.~Bal{\'a}zsi, A.~Van~Oudenaarden and J.J.~Collins, {Cellular decision making
  and biological noise: from microbes to mammals},  {Cell} {\bfseries 144}
  (2011) 910.

\bibitem{li2014linear}
Y.~Li, M.~Yi and X.~Zou, {The linear interplay of intrinsic and extrinsic
  noises ensures a high accuracy of cell fate selection in budding yeast},
  {Scientific reports} {\bfseries 4} (2014) 5764.

\bibitem{han2008understanding}
J.-D.J.~Han, {Understanding biological functions through molecular networks},
  {Cell research} {\bfseries 18} (2008) 224.

\bibitem{ackermann2015functional}
M.~Ackermann, {A functional perspective on phenotypic heterogeneity in
  microorganisms},  {Nature Reviews Microbiology} {\bfseries 13} (2015) 497.

\bibitem{del2002periodicity}
C.A.~Del~Negro, C.G.~Wilson, R.J.~Butera, H.~Rigatto and J.C.~Smith,
  {Periodicity, mixed-mode oscillations, and quasiperiodicityin a
  rhythm-generating neural network},  {Biophysical Journal} {\bfseries 82}
  (2002) 206.

\bibitem{houart1999bursting}
G.~Houart, G.~Dupont and A.~Goldbeter, {Bursting, chaos and birhythmicity
  originating from self-modulation of the inositol 1, 4, 5-trisphosphate signal
  in a model for intracellular ca2+ oscillations},  {Bulletin of mathematical
  biology} {\bfseries 61} (1999) 507.

\bibitem{matsu2006cytosolic}
T.~Matsu-ura, T.~Michikawa, T.~Inoue, A.~Miyawaki, M.~Yoshida and K.~Mikoshiba,
  {Cytosolic inositol 1, 4, 5-trisphosphate dynamics during intracellular
  calcium oscillations in living cells},  {The Journal of cell biology}
  {\bfseries 173} (2006) 755.

\bibitem{perc2009prevalence}
M.~Perc, M.~Rupnik, M.~Gosak and M.~Marhl, {Prevalence of stochasticity in
  experimentally observed responses of pancreatic acinar cells to
  acetylcholine},  {Chaos: An Interdisciplinary Journal of Nonlinear Science}
  {\bfseries 19} (2009).

\bibitem{tamarina2005inositol}
N.A.~Tamarina, A.~Kuznetsov, C.J.~Rhodes, V.P.~Bindokas and L.H.~Philipson,
  {Inositol (1, 4, 5)-trisphosphate dynamics and intracellular calcium
  oscillations in pancreatic $\beta$-cells},  {Diabetes} {\bfseries 54} (2005)
  3073.

\bibitem{wu2005phase}
D.~Wu, Y.~Jia, L.~Yang, Q.~Liu and X.~Zhan, {Phase synchronization and
  coherence resonance of stochastic calcium oscillations in coupled
  hepatocytes},  {Biophysical chemistry} {\bfseries 115} (2005) 37.

\bibitem{collier2000calcium}
M.~Collier, G.~Ji, Y.-X.~Wang and M.~Kotlikoff, {Calcium-induced calcium
  release in smooth muscle: loose coupling between the action potential and
  calcium release},  {The Journal of general physiology} {\bfseries 115} (2000)
  653.

\bibitem{meng2007calcium}
F.~Meng, W.~To, J.~Kirkman-Brown, P.~Kumar and Y.~Gu, {Calcium oscillations
  induced by atp in human umbilical cord smooth muscle cells},  {Journal of
  cellular physiology} {\bfseries 213} (2007) 79.

\bibitem{verkhratsky1996calcium}
A.~Verkhratsky and A.~Shmigol, {Calcium-induced calcium release in neurones},
  {Cell calcium} {\bfseries 19} (1996) 1.

\bibitem{berridge1994spatial}
M.J.~Berridge and G.~Dupont, {Spatial and temporal signalling by calcium},
  {Current opinion in cell biology} {\bfseries 6} (1994) 267.

\bibitem{thurley2012fundamental}
K.~Thurley, A.~Skupin, R.~Thul and M.~Falcke, {Fundamental properties of ca2+
  signals},  {Biochimica et Biophysica Acta-General Subjects} {\bfseries 1820}
  (2012) 1185.

\bibitem{dolmetsch1998calcium}
R.E.~Dolmetsch, K.~Xu and R.S.~Lewis, {Calcium oscillations increase the
  efficiency and specificity of gene expression},  {Nature} {\bfseries 392}
  (1998) 933.

\bibitem{humeau2018calcium}
J.~Humeau, J.M.~Bravo-San~Pedro, I.~Vitale, L.~Nunez, C.~Villalobos, G.~Kroemer
  et~al., {Calcium signaling and cell cycle:progression or death},  {Cell
  calcium} {\bfseries 70} (2018) 3.

\bibitem{pinto2016studying}
M.C.~Pinto, F.M.~Tonelli, A.L.~Vieira, A.H.~Kihara, H.~Ulrich and R.R.~Resende,
  {Studying complex system: calcium oscillations as attractor of cell
  differentiation},  {Integrative Biology} {\bfseries 8} (2016) 130.

\bibitem{xu2012potential}
L.~Xu, F.~Zhang, E.~Wang and J.~Wang, {The potential and flux landscape,
  lyapunov function and non-equilibrium thermodynamics for dynamic systems and
  networks with an application to signal-induced ca2+ oscillation},
  {Nonlinearity} {\bfseries 26} (2012) R69.

\bibitem{puebla2005controlling}
H.~Puebla, {Controlling intracellular calcium oscillations and waves},
  {Journal of Biological Systems} {\bfseries 13} (2005) 173.

\bibitem{borghans1997complex}
J.M.~Borghans, G.~Dupont and A.~Goldbeter, {Complex intracellular calcium
  oscillations a theoretical exploration of possible mechanisms},  {Biophysical
  chemistry} {\bfseries 66} (1997) 25.

\bibitem{Terrar2020}
D.A.~Terrar, {Calcium signaling in the heart},  in {Calcium Signaling},
  M.S.~Islam, ed., (Cham), pp.~395--443, Springer International Publishing
  (2020), \href{https://doi.org/10.1007/978-3-030-12457-1_16}{DOI}.

\bibitem{perc2007periodic}
M.~Perc, M.~Gosak and M.~Marhl, {Periodic calcium waves in coupled cells
  induced by internal noise},  {Chemical Physics Letters} {\bfseries 437}
  (2007) 143.

\bibitem{folz2021interplay}
F.~Folz, K.~Mehlhorn and G.~Morigi, {Interplay of periodic dynamics and noise:
  Insights from a simple adaptive system},  {Physical Review E} {\bfseries 104}
  (2021) 054215.

\bibitem{friedrich2011approaching}
R.~Friedrich, J.~Peinke, M.~Sahimi and M.R.R.~Tabar, {Approaching complexity by
  stochastic methods: From biological systems to turbulence},  {Physics
  Reports} {\bfseries 506} (2011) 87.

\bibitem{roli2019complexity}
A.~Roli, A.~Ligot and M.~Birattari, {Complexity measures: open questions and
  novel opportunities in the automatic design and analysis of robot swarms},
  {Frontiers in Robotics and AI} {\bfseries 6} (2019) 130.

\bibitem{lloyd2001measures}
S.~Lloyd, {Measures of complexity: a nonexhaustive list},  {IEEE Control
  Systems Magazine} {\bfseries 21} (2001) 7.

\bibitem{grassberger1991information}
P.~Grassberger, {Information and complexity measures in dynamical systems},  in
  {Information dynamics}, pp.~15--33, Springer (1991).

\bibitem{shannon1948mathematical}
C.E.~Shannon, {A mathematical theory of communication},  {The Bell system
  technical journal} {\bfseries 27} (1948) 379.

\bibitem{bandt2002permutation}
C.~Bandt and B.~Pompe, {Permutation entropy: a natural complexity measure for
  time series},  {Physical review letters} {\bfseries 88} (2002) 174102.

\bibitem{lopez1995statistical}
R.~Lopez-Ruiz, H.L.~Mancini and X.~Calbet, {A statistical measure of
  complexity},  {Physics letters A} {\bfseries 209} (1995) 321.

\bibitem{rosso2007distinguishing}
O.A.~Rosso, H.~Larrondo, M.T.~Martin, A.~Plastino and M.A.~Fuentes,
  {Distinguishing noise from chaos},  {Physical review letters} {\bfseries 99}
  (2007) 154102.

\bibitem{gillespie1977exact}
D.T.Gillespie, {Exact stochastic simulation of coupled chemical reactions},
  {The journal of physical chemistry} {\bfseries 81} (1977) 2340.

\bibitem{gillespie1996multivariate}
D.T.~Gillespie, {The multivariate langevin and fokker--planck equations},
  {American Journal of Physics} {\bfseries 64} (1996) 1246.

\bibitem{green1993adenine}
A.K.~Green, C.J.~Dixon, A.G.~McLennan, P.H.~Cobbold and M.J.~Fisher, {Adenine
  dinucleotide-mediated cytosolic free ca2+ oscillations in single
  hepatocytes},  {FEBS letters} {\bfseries 322} (1993) 197.

\bibitem{marrero1994taurolithocholate}
I.~Marrero, A.~Sanchez-Bueno, P.~Cobbold and C.~Dixon, {Taurolithocholate and
  taurolithocholate 3-sulphate exert different effects on cytosolic free ca2+
  concentration in rat hepatocytes},  \href{https://doi.org/10.1042/bj3000383}
  {Biochemical Journal} {\bfseries 300} (1994) 383.

\bibitem{pessa2021ordpy}
A.A.Pessa and H.V.~Ribeiro, {ordpy: A python package for data analysis with
  permutation entropy and ordinal network methods},  {Chaos: An
  Interdisciplinary Journal of Nonlinear Science} {\bfseries 31} (2021).

\bibitem{riedl2013practical}
M.Riedl, A.~M{\"u}ller and N.~Wessel, {Practical considerations of permutation
  entropy: A tutorial review},  {The European Physical Journal Special Topics}
  {\bfseries 222} (2013) 249.

\bibitem{smaal2021complexity}
N.~Smaal and J.R.C.~Piqueira, {Complexity measures for maxwell--boltzmann
  distribution},  {Complexity} {\bfseries 2021} (2021) 1.

\bibitem{martin2006generalized}
M.~Martin, A.~Plastino and O.~Rosso, {Generalized statistical complexity
  measures: Geometrical and analytical properties},  {Physica A: Statistical
  Mechanics and its Applications} {\bfseries 369} (2006) 439.

\bibitem{kowalski2005entropic}
A.~Kowalski, M.~Martin, A.~Plastino and O.~Rosso, {Entropic non-triviality, the
  classical limit and geometry-dynamics correlations},  {International Journal
  of Modern Physics B} {\bfseries 19} (2005) 2273.

\bibitem{zunino2007characterization}
L.~Zunino, D.~P{\'e}rez, M.~Mart{\'\i}n, A.~Plastino, M.~Garavaglia and
  O.~Rosso, {Characterization of gaussian self-similar stochastic processes
  using wavelet-based informational tools},  {Physical Review E} {\bfseries 75}
  (2007) 021115.

\bibitem{zhu2007mesoscopic}
C.-l.~Zhu, Y.~Jia, Q.~Liu, L.-j.~Yang and X.~Zhan, {A mesoscopic stochastic
  mechanism of cytosolic calcium oscillations},  {Biophysical Chemistry}
  {\bfseries 125} (2007) 201.

\bibitem{BLANCO2017547}
A.~Blanco and G.~Blanco, {Chapter 25 - biochemical basis of endocrinology (i)
  receptors and signal transduction},  in {Medical Biochemistry}, A.~Blanco and
  G.~Blanco, eds., pp.~547--572, Academic Press (2017),
  \href{https://doi.org/https://doi.org/10.1016/B978-0-12-803550-4.00025-2}{DOI}.

\bibitem{WEAVER2020321}
C.M.~Weaver, {Chapter 19 - calcium},  in {Present Knowledge in Nutrition
  (Eleventh Edition)}, B.P.~Marriott, D.F.~Birt, V.A.~Stallings and A.A.~Yates,
  eds., pp.~321--334, Academic Press (2020),
  \href{https://doi.org/https://doi.org/10.1016/B978-0-323-66162-1.00019-6}{DOI}.

\bibitem{murphy255rich}
E.~Murphy and K.~Coll, {Rich. tl \& williamson, jr (1980) j},  {Biol. Chem}
  {\bfseries 255} 6600.

\bibitem{williamson1983cytosolic}
J.~Williamson, R.J.~Williams, K.~Coll and A.~Thomas, {Cytosolic free ca2+
  concentration and intracellular calcium distribution of ca2+-tolerant
  isolated heart cells.},  {Journal of Biological Chemistry} {\bfseries 258}
  (1983) 13411.

\bibitem{gillespie2002chemical}
D.T.Gillespie, {The chemical langevin and fokker- planck equations for the
  reversible isomerization reaction},  {The Journal of Physical Chemistry A}
  {\bfseries 106} (2002) 5063.

\bibitem{thounaojam2022stochastic}
U.S.~Thounaojam, {Stochastic chaos in chemical lorenz system: Interplay of
  intrinsic noise and nonlinearity},  {Chaos, Solitons \& Fractals} {\bfseries
  165} (2022) 112763.

\bibitem{pomeau1984order}
Y.~Pomeau and C.~Vidal, {Order within chaos: towards a deterministic approach
  to turbulence}, Wiley (1984).

\bibitem{wu1993internal}
X.-G.~Wu and R.~Kapral, {Internal fluctuations and deterministic chemical
  chaos},  {PRL} {\bfseries 70} (1993) 1940.

\bibitem{willamowski1980irregular}
K.-D.~Willamowski and O.~R{\"o}ssler, {Irregular oscillations in a realistic
  abstract quadratic mass action system},  {Zeitschrift f{\"u}r Naturforschung
  A} {\bfseries 35} (1980) 317.

\bibitem{calbet2001tendency}
X.~Calbet and R.~L{\'o}pez-Ruiz, {Tendency towards maximum complexity in a
  nonequilibrium isolated system},  {Physical Review E} {\bfseries 63} (2001)
  066116.

\bibitem{jeon2010fractional}
J.-H.~Jeon and R.~Metzler, {Fractional brownian motion and motion governed by
  the fractional langevin equation in confined geometries},  {Physical Review
  E} {\bfseries 81} (2010) 021103.

\bibitem{hong2006power}
S.~Hong, J.W.~Bodfish and K.M.~Newell, {Power-law scaling for macroscopic
  entropy and microscopic complexity: evidence from human movement and
  posture},  {Chaos: An Interdisciplinary Journal of Nonlinear Science}
  {\bfseries 16} (2006).

\bibitem{mehri2016power}
A.~Mehri and S.M.~Lashkari, {Power-law regularities in human language},  {The
  European Physical Journal B} {\bfseries 89} (2016) 1.

\bibitem{nogueira2017exploring}
M.~Nogueira, {Exploring the link between multiscale entropy and fractal scaling
  behavior in near-surface wind},  {PloS one} {\bfseries 12} (2017) e0173994.

\bibitem{kantelhardt2001detecting}
J.~Kantelhardt, E.~Koscielny-Bunde, H.H.~Rego, S.~Havlin and A.~Bunde,
  {Detecting long-range correlations with detrended fluctuation analysis},
  {Phys. A: Stat.} {\bfseries 295} (2001) 441.

\bibitem{kantelhardt2002multifractal}
J.W.~Kantelhardt, S.A.~Zschiegner, E.~Koscielny-Bunde, S.~Havlin, A.~Bunde and
  H.E.~Stanley, {Multifractal detrended fluctuation analysis of nonstationary
  time series},  {Phys. A: Stat.} {\bfseries 316} (2002) 87.

\bibitem{peng1994mosaic}
C.-K.~Peng, S.V.~Buldyrev, S.~Havlin, M.~Simons, H.E.~Stanley and
  A.L.~Goldberger, {Mosaic organization of dna nucleotides},  {Physical review
  e} {\bfseries 49} (1994) 1685.

\bibitem{zhou2006inverse}
W.-X.~Zhou, D.~Sornette and W.-K.~Yuan, {Inverse statistics and multifractality
  of exit distances in 3d fully developed turbulence},  {Physica D: Nonlinear
  Phenomena} {\bfseries 214} (2006) 55.

\bibitem{li2005internal}
H.~Li, Z.~Hou and H.~Xin, {Internal noise stochastic resonance for
  intracellular calcium oscillations in a cell system},  {Physical Review E}
  {\bfseries 71} (2005) 061916.

\bibitem{li2005internall}
H.~Li, Z.~Hou and H.~Xin, {Internal noise enhanced detection of hormonal signal
  through intracellular calcium oscillations},  {Chemical physics letters}
  {\bfseries 402} 444.

\bibitem{boccaletti2006complex}
S.Boccaletti, V.~Latora, Y.~Moreno, M.~Chavez and D.-U.~Hwang, {Complex
  networks: Structure and dynamics},  {Physics reports} {\bfseries 424} (2006)
  175.

\bibitem{eskov2016evolution}
V.~Eskov, V.~Eskov, J.~Vochmina and T.~Gavrilenko, {The evolution of the
  chaotic dynamics of collective modes as a method for the behavioral
  description of living systems},  {Moscow university physics bulletin}
  {\bfseries 71} (2016) 143.

\bibitem{korn2003there}
H.~Korn and P.~Faure, {Is there chaos in the brain? ii. experimental evidence
  and related models},  {Comptes rendus biologies} {\bfseries 326} (2003) 787.

\bibitem{riaz2008chaotic}
A.~Riaz and M.~Ali, {Chaotic communications, their applications and advantages
  over traditional methods of communication},  in {2008 6th International
  Symposium on Communication Systems, Networks and Digital Signal Processing},
  pp.~21--24, IEEE, 2008.

\bibitem{seara2021irreversibility}
D.S.~Seara, B.B.~Machta and M.P.~Murrell, {Irreversibility in dynamical phases
  and transitions},  {Nature communications} {\bfseries 12} (2021) 392.

\bibitem{seifert2004fluctuation}
U.~Seifert, {Fluctuation theorem for birth--death or chemical master equations
  with time-dependent rates},  {J. Phys. A} {\bfseries 37} (2004) L517.

\bibitem{andrieux2007entropy}
D.~Andrieux, P.~Gaspard, S.~Ciliberto, N.~Garnier, S.~Joubaud and A.~Petrosyan,
  {Entropy production and time asymmetry in nonequilibrium fluctuations},
  {Physical review letters} {\bfseries 98} (2007) 150601.

\bibitem{xiao2009stochastic}
T.~Xiao, Z.~Hou and H.~Xin, {Stochastic thermodynamics in mesoscopic chemical
  oscillation systems},  {The Journal of Physical Chemistry B} {\bfseries 113}
  (2009) 9316.

\bibitem{RePEc:cwl:cwldpp:1164}
B.~Mandelbrot, A.~Fisher and L.~Calvet, {{A Multifractal Model of Asset
  Returns}},  Cowles Foundation Discussion Papers
  \href{https://ideas.repec.org/p/cwl/cwldpp/1164.html}{1164}, Cowles
  Foundation for Research in Economics, Yale University (Sept., 1997).

\end{thebibliography}\endgroup
\end{document}